\documentclass[twocolumn,pra,twocolumn,superscriptaddress]{revtex4}
\usepackage{color}
\usepackage{amsmath}
\usepackage{amssymb}
\usepackage{graphicx}
\usepackage{tikz}
\usepackage{multirow}
\usepackage{float}
\usepackage{bbm}
\usepackage{dsfont}
\usepackage{cases}
\usepackage{mathtools}

\usepackage{epsfig}
\usepackage{verbatim}
\usepackage{array}
\usepackage{setspace}

\usepackage{scalerel}
\usepackage{stackengine,wasysym}
\usepackage{empheq, nccmath}

\usepackage[T1]{fontenc}

\usepackage[unicode=true,
bookmarks=true,bookmarksnumbered=false,bookmarksopen=false,
breaklinks=false,pdfborder={0 0 1},backref=false,colorlinks=true]
{hyperref}
\hypersetup{
	linkcolor=magenta, urlcolor=blue, citecolor=blue, pdfstartview={FitH}, hyperfootnotes=false, unicode=true}

\makeatletter
\@ifundefined{textcolor}{}
{%
	\definecolor{BLACK}{gray}{0}
	\definecolor{WHITE}{gray}{1}
	\definecolor{RED}{rgb}{1,0,0}
	\definecolor{GREEN}{rgb}{0,1,0}
	\definecolor{BLUE}{rgb}{0,0,1}
	\definecolor{CYAN}{cmyk}{1,0,0,0}
	\definecolor{MAGENTA}{cmyk}{0,1,0,0}
	\definecolor{YELLOW}{cmyk}{0,0,1,0}
}

\newsavebox\myboxA
\newsavebox\myboxB
\newlength\mylenA

\newcommand*\xoverline[2][0.75]{%
	\sbox{\myboxA}{$\m@th#2$}%
	\setbox\myboxB\null
	\ht\myboxB=\ht\myboxA%
	\dp\myboxB=\dp\myboxA%
	\wd\myboxB=#1\wd\myboxA
	\sbox\myboxB{$\m@th\overline{\copy\myboxB}$}
	\setlength\mylenA{\the\wd\myboxA}
	\addtolength\mylenA{-\the\wd\myboxB}%
	\ifdim\wd\myboxB<\wd\myboxA%
	\rlap{\hskip 0.5\mylenA\usebox\myboxB}{\usebox\myboxA}%
	\else
	\hskip -0.5\mylenA\rlap{\usebox\myboxA}{\hskip 0.5\mylenA\usebox\myboxB}%
	\fi}

\usepackage{amsfonts}\usepackage{tabularx}\usepackage{dcolumn}\usepackage{bm}\usepackage{graphicx}\usepackage{epstopdf}

\hypersetup{urlcolor=blue}

\let\save@mathaccent\mathaccent
\newcommand*\if@single[3]{%
	\setbox0\hbox{${\mathaccent"0362{#1}}^H$}%
	\setbox2\hbox{${\mathaccent"0362{\kern0pt#1}}^H$}%
	\ifdim\ht0=\ht2 #3\else #2\fi
}
\newcommand*\rel@kern[1]{\kern#1\dimexpr\macc@kerna}
\newcommand*\wideaccent[2]{\@ifnextchar^{{\wide@accent{#1}{#2}{0}}}{\wide@accent{#1}{#2}{1}}}
\newcommand*\wide@accent[3]{\if@single{#2}{\wide@accent@{#1}{#2}{#3}{1}}{\wide@accent@{#1}{#2}{#3}{2}}}
\newcommand*\wide@accent@[4]{%
	\begingroup
	\def\mathaccent##1##2{%
		\let\mathaccent\save@mathaccent
		\if#42 \let\macc@nucleus\first@char \fi
		\setbox\z@\hbox{$\macc@style{\macc@nucleus}_{}$}%
		\setbox\tw@\hbox{$\macc@style{\macc@nucleus}{}_{}$}%
		\dimen@\wd\tw@
		\advance\dimen@-\wd\z@
		\divide\dimen@ 3
		\@tempdima\wd\tw@
		\advance\@tempdima-\scriptspace
		\divide\@tempdima 10
		\advance\dimen@-\@tempdima
		\ifdim\dimen@>\z@ \dimen@0pt\fi
		\rel@kern{0.6}\kern-\dimen@
		\if#41
		#1{\rel@kern{-0.6}\kern\dimen@\macc@nucleus\rel@kern{0.4}\kern\dimen@}%
		\advance\dimen@0.4\dimexpr\macc@kerna
		\let\final@kern#3%
		\ifdim\dimen@<\z@ \let\final@kern1\fi
		\if\final@kern1 \kern-\dimen@\fi
		\else
		#1{\rel@kern{-0.6}\kern\dimen@#2}%
		\fi
	}%
	\macc@depth\@ne
	\let\math@bgroup\@empty \let\math@egroup\macc@set@skewchar
	\mathsurround\z@ \frozen@everymath{\mathgroup\macc@group\relax}%
	\macc@set@skewchar\relax
	\let\mathaccentV\macc@nested@a
	\if#41
	\macc@nested@a\relax111{#2}%
	\else
	\def\gobble@till@marker##1\endmarker{}%
	\futurelet\first@char\gobble@till@marker#2\endmarker
	\ifcat\noexpand\first@char A\else
	\def\first@char{}%
	\fi
	\macc@nested@a\relax111{\first@char}%
	\fi
	\endgroup
}

\makeatother

\allowdisplaybreaks
\newcolumntype{C}[1]{>{\centering\arraybackslash$}p{#1}<{$}}

\makeatletter
\let\alignts@preamble\align@preamble
\patchcmd{\alignts@preamble}{\displaystyle}{\textstyle}{}{}
\patchcmd{\alignts@preamble}{\displaystyle}{\textstyle}{}{}

\def\alignts{\let\align@preamble\alignts@preamble\start@align\@ne\st@rredfalse\m@ne}

\makeatother

\begin{document}
	
	\title{Universal control of superexchange in linear triple quantum dots with an empty mediator}
	
	\author{Guo Xuan Chan}
	\affiliation{Department of Physics, City University of Hong Kong, Tat Chee Avenue, Kowloon, Hong Kong SAR, P. R. China, and City University of Hong Kong Shenzhen Research Institute, Shenzhen, Guangdong 518057, P. R. China}
	
	\author{Peihao Huang}
	\email{huangph@sustech.edu.cn}
	\affiliation{Shenzhen Institute for Quantum Science and Engineering, Southern University of Science and Technology, Shenzhen, Guangdong 518055, P. R. China}
	\affiliation{International Quantum Academy, Shenzhen, Guangdong 518048, P. R. China}
	\affiliation{Guangdong Provincial Key Laboratory of Quantum Science and Engineering, Southern University of Science and Technology, Shenzhen, Guangdong 518055, P. R. China}
	
	\author{Xin Wang}
	\email{x.wang@cityu.edu.hk}
	\affiliation{Department of Physics, City University of Hong Kong, Tat Chee Avenue, Kowloon, Hong Kong SAR, P. R. China, and City University of Hong Kong Shenzhen Research Institute, Shenzhen, Guangdong 518057, P. R. China}
	
	\date{\today}
	
	\begin{abstract}
	Superexchange is one of the vital resources to realize long-range interaction between distant spins for large-scale quantum computing. Recent experiments have demonstrated coherent oscillations between logical states defined by remote spins whose coupling is given by the superexchange interaction mediated by central spins. Excavating the potential of superexchange requires a full understanding of the interaction in terms of control parameters, which is still lacking in literature. Here, using full configuration interaction calculations, we study a two-electron system in a linear triple-quantum-dot device in which the left and right dots are occupied by a single electron each, whose spin states are defined as qubits. The numerical nature of the full configuration interaction calculations allows access to the microscopic details of the quantum-dot confining potential and electronic wavefunctions, some of which are overlooked in the celebrated Hubbard model but turn out to be critical for the behavior of superexchange. Following experimental demonstrations of superexchange interactions, we focus on the detuning regime where the charge ground state yields an empty middle dot. We have found that, when the detunings at the left and right dots are leveled, the superexchange can exhibit a non-monotonic behavior which ranges from positive to negative values as a function of the middle-dot detuning. We further show that a larger relative detuning between the left and right dots causes the magnitude of the superexchange to increase (decrease) for an originally positive (negative) superexchange. We then proceed to show the results for a much larger left-right dot detuning. Using a Hubbard-like model, we present analytical expressions of the superexchange and have found that they conform well qualitatively with the numerical results. Our results suggest that even a simple configuration of delocalized two-electron states in a linear triple-quantum-dot device exhibits superexchange energy with non-trivial behaviors, which could have important applications in spin-based quantum computing.
	
	\end{abstract}

	\maketitle
	\section{Introduction}
	The Heisenberg exchange interaction lays the foundation of spin-based quantum computation in semiconductor quantum-dots \cite{Loss.98,DiVincenzo.00,Petta.05,Foletti.09,Laird.10,Nowack.11,Medford.13,Shi.12,Kim.14,Kevin.15,Shim.16,Sala.17,Russ.18,Zajac.18,Watson.18,Gullans.19,Sigillito.19,Hendrickx.21,Philips.22}. Utilizing the electrostatic nature of the electron configurations in quantum-dot arrays, exchange-based quantum gates offer tunability of the interaction strength, permitting universal quantum computation. However, exchange coupling only materializes for nearest-neighbor interactions, limiting the realization of large-scale quantum computing in which long-range interactions are desired. 
	
	To achieve scalable quantum computation, various coupling schemes for long-range interactions have been proposed, including capacitive Coulomb interaction \cite{Taylor.05,Shulman.12,Pal.14,Pal.15,Li.15,Nichol.17,Wolfe.17,Buterakos.18,Neyens.19,ChanGX.21,Huang.21}, hybrid spin-cQED (circuit quantum electrodynamics) architectures which utilize couplings between electron spins and photons in microwave cavities \cite{Huang.13,Mi.17,Mi.17.2,Mi.18,Woerkom.18,Koski.20,Borjans.20,Burkard.20,Borjans.20.2,Kratochwil.21,Ruskov.21} and the method of electron shuttling in which remote spins are physically brought closer to enable nearest-neighbor exchange interaction \cite{Nakajima.18,Fujita.17,Mills.19,Buonacorsi.20,Ginzel.20,Jadot.21,Noiri.22}. The former two schemes suffer from susceptibility to charge noises due to the introduced dipole for the enhancement of the coupling strength while the latter requires a careful pulse design to preserve spin and phase coherence during the shuttling operation \cite{Mills.19,Noiri.22}.
	
	An alternative coupling scheme involves mediators formed by electron spin states between distant logical spins, termed as the superexchange \cite{Bose.03,Wojcik.05,Campos.06,Bose.07,Friesen.07,Oh.10,Oh.11,Braakman.13,Sanchez.14,Sanchez.14.2,Busl.13,Rancic.17,Baart.17,Croot.18,Malinowski.19,Qiao.20,Deng.20,Chan.21,Qiao.21,Knorzer.22}. Virtual exchange through mediators enables a long-range linkage between remote spins, giving rise to the superexchange interaction. Current experimental demonstrations on superexchange include single-quantum-dot mediators with various electron numbers (e.g.~zero  \cite{Baart.17}, one \cite{Chan.21} or multiple electrons \cite{Malinowski.19}) and singly-occupied quantum-dot chain \cite{Qiao.21}. Also, the spatial separation of logical spin states can potentially serve to mitigate crosstalk during the single-qubit operation \cite{Croot.18}. 
	
For a linear triple-quantum-dot (TQD) device with two electrons, systematic studies on the superexchange interaction between two spins in the singly-occupied outmost dots have been conducted \cite{Deng.20,Baart.17,Rancic.17}. In particular, Ref.~\onlinecite{Baart.17} has concluded that the superexchange energy $J_\text{se}$ is positive and monotonic with respect to the relative detuning between outmost dots. Similarly, based on the results derived from Hubbard model, Ref.~\onlinecite{Rancic.17} has theoretically demonstrated that $J_\text{se}$ is always positive in the detuning region where the electron occupation in the middle dot is zero.
In another theoretical work on superexchange in a triangular TQD, Ref.~\onlinecite{Deng.20} shows that the presence of an empty quantum-dot mediator contributes a modest enhancement to the nearest-neighbor exchange between two logical spins.

	Although the aforementioned works are instrumental in understanding the mechanism of superexchange in quantum-dot devices, more comprehensive studies exploring the superexchange interaction in other regimes of parameters are still lacking. In this work, we study superexchange interaction in a linear TQD device in which the logical states are defined by two electron spins in the left and right dots while maintaining an empty middle dot. Following the experimental demonstrations in quantum-dot devices \cite{Malinowski.19,Chan.21,Qiao.21}, we only consider the detuning region where the charge ground state yields an empty middle dot. In particular, we study the tunability of $J_\text{se}$ in terms of the detuning values and the existence of sweet spots, which are crucial for high-fidelity gate operations \cite{Hu.06,Koch.07,Stopa.08,Culcer.09,Li.10,Nielsen.10,Yang.11,Medford.13,Medford.13.2,Taylor.13,Russ.15,Reed.16,Martins.16}.
	
	We employ full Configuration Interaction (full CI) calculations to numerically simulate superexchange energy $J_\text{se}$. Using the full CI method allows us to associate the behaviors of $J_\text{se}$ to the details of the dot parameters, i.e.~confinement strengths and inter-dot distances. Moreover, adopting full CI techniques is important to avoid obtaining erroneous exchange couplings between closely-spaced quantum-dots (QDs), a common issue encountered by the minimal basis models, i.e.~Heitler-London (HL) and Hund-Mulliken (HM) approximations \cite{Li.10,White.18}. It should be noted that, although CI techniques have been adopted in Ref.~\onlinecite{Deng.20} for a linear TQD, the results are limited to $J_\text{se}$ of a two-electron mediator at a particular detuning value. In contrast, our work focuses on $J_\text{se}$ of an empty mediator dot for ranges of detuning values.
	
	We will show that, when the left-right detuning $\varepsilon$ is zero, $J_\text{se}$ yields a non-monotonic behavior which switches from positive to negative values as a function of the middle-dot detuning $\Delta$. The switching sign of $J_\text{se}$ is found to be present only for a larger quantum-dot confinement strength with a smaller inter-dot distance. Also, the non-monotonicity of $J_\text{se}$ gives rise to a sweet spot with respect to both the middle dot detuning and left-right detuning.
	We will also show that when a sufficiently large left-right detuning $\varepsilon$ is applied, an originally negative $J_\text{se}$ switches to a positive value. We are able to understand the sign switching of $J_\text{se}$ using a generic Hubbard model, in which the Coulomb exchange terms are included as explained below.
	In the charge regime where the middle dot is empty, we confirm that the sign-switching and sweet spot of $J_\text{se}$ cannot be reproduced using Hubbard Model \cite{Rancic.17} or the extended Hubbard Model \cite{Kotlyar.98}, highlighting the significance of the full CI method.
Our results suggest that the superexchange, among other novel coupling schemes in multi-quantum-dot devices, may open a viable route for scalable quantum-dot quantum information processing.

	\begin{figure}[tb]
		\includegraphics[width=0.95\linewidth]{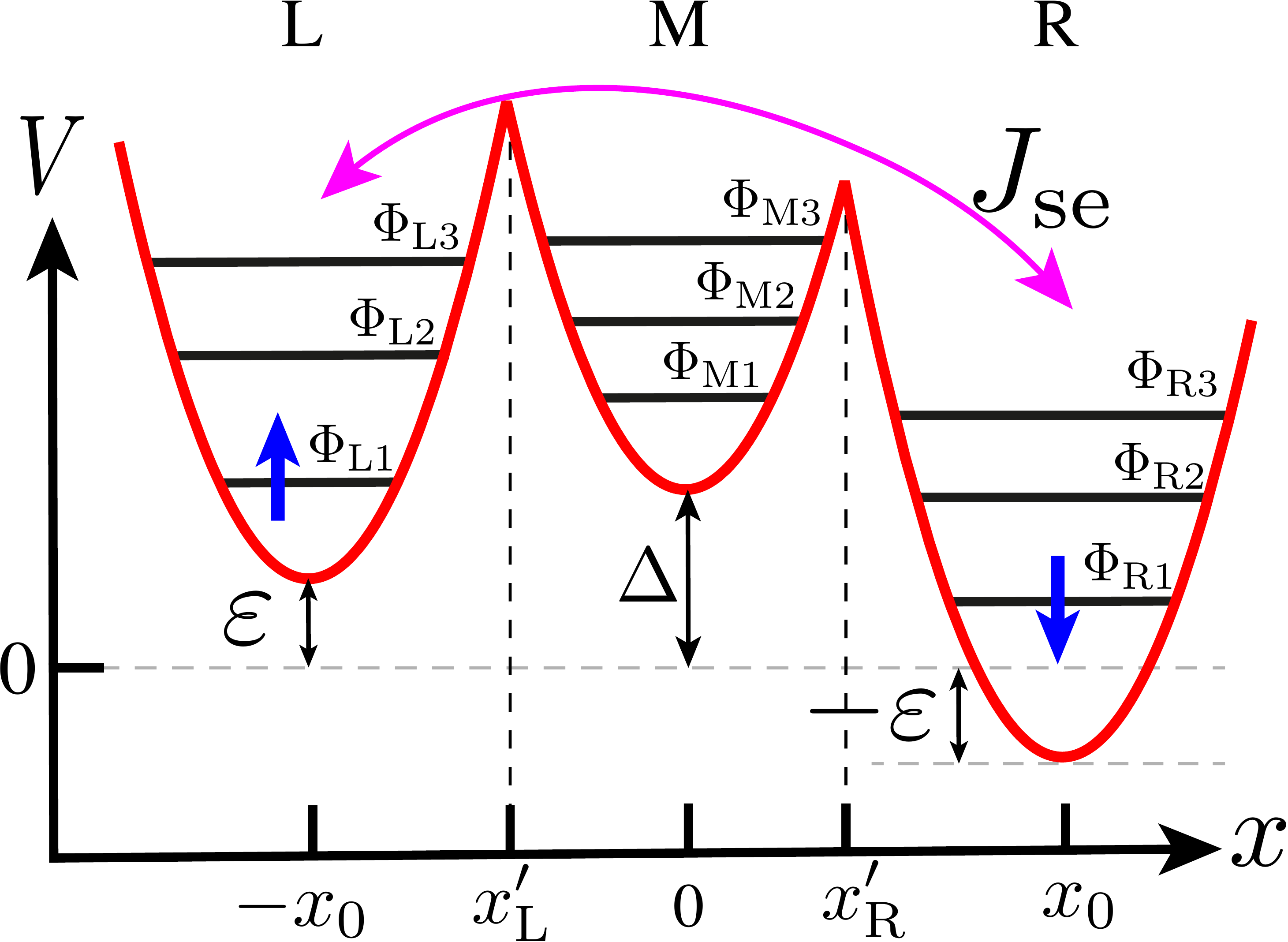}
		\caption{Schematic illustration of a TQD device. $\Phi_{j\alpha}$ indicates the $\alpha$-th orthogonalized Fock-Darwin (F-D) states in dot $j$, where $j \in \{\text{L}, \text{M}, \text{R}\}$. $J_{\text{se}}$ (magenta arrow) indicates the superexchange between dot L and R.}
		\label{fig:V}
	\end{figure}
		
	This paper is organized as follows. In Sec.~\ref{sec:model}, we provide the details of the TQD system of interest, including the full CI calculations to obtain the relevant eigenvalues and eigenstates (Sec.~\ref{subsec:CI}) and the analytical expressions of $J_\text{se}$ (Sec.~\ref{subsec:Jse}). In Sec.~\ref{sec:result}, we present the numerical results of $J_\text{se}$. In Sec.~\ref{sec:JseVSlrmD}, we show the extensive results of $J_\text{se}$, i.e.~the values of $J_\text{se}$ as functions of $\Delta$ and $\varepsilon$, and identify three main behaviors of $J_\text{se}$. In Secs.~\ref{sec:JseDeltaLR0}-\ref{subsubsec:JswitchDLRLarge}, we discuss those behaviors separately: (1) Sec.~\ref{sec:JseDeltaLR0} discusses the behavior of $J_\text{se}$ as a function of $\Delta$ when $\varepsilon=0$; (2) Sec.~\ref{subsubsec:JseS002lS011in101} discusses the changes of $J_\text{se}$ for moderate values of $\varepsilon$, i.e.~$\varepsilon>0$; (3) Sec.~\ref{subsubsec:JswitchDLRLarge} discusses  $J_\text{se}$ for a large $\varepsilon$, i.e.~$\varepsilon\gg0$. We then compare our work with other relevant works in Sec.~\ref{sec:compareOther}. We conclude our findings in Sec.~\ref{sec:conclusion}.
	
	\section{Model}\label{sec:model}

	\begin{figure*}[t]
		\includegraphics[width=\linewidth]{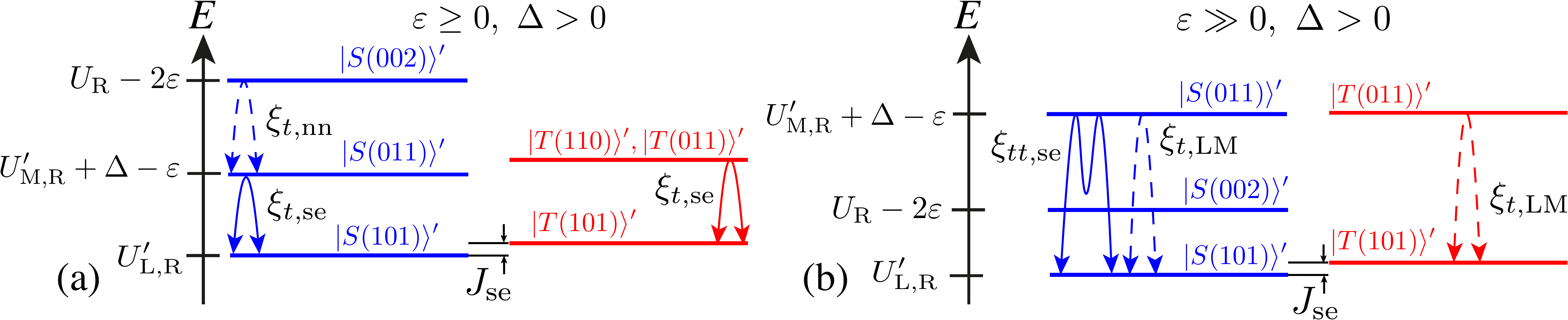}
		\caption{Schematic energy level structures of the system for (a) $\varepsilon\geq0$ and (b) $\varepsilon\gg0$. Note, since the energy splittings between states vary depending on $\Delta$ and $\varepsilon$, the energy levels are not to scale.}
		\label{fig:lvlDia}
	\end{figure*}

	\subsection{Configuration Interaction (CI)}\label{subsec:CI}

	We consider an $n$-electron system Hamiltonian
	\begin{equation}\label{eq:H}
		H=\sum_{j=1}^n h_j + H_C,
	\end{equation}
	with the single-particle Hamiltonian $h_j = {(-i\hbar \nabla_j+e \mathbf{A}/c)^2}/{2m^*}+V(\mathbf{r})+g^*\mu_B \mathbf{B}\cdot \mathbf{S}$ and the Coulomb interaction $H_C = \sum{e^2}/\epsilon\left|\mathbf{r}_j-\mathbf{r}_k\right|$, where $\mathbf{A}$ is the vector potential and $\mathbf{B}$ is the magnetic field. As shown schematically in Fig.~\ref{fig:V}, the potential function of a triple-quantum-dot (TQD), $V(\mathbf{r})$, is modeled as
	
	\begin{equation}\label{eq:V}
		V(\mathbf{r})=
		\begin{cases}
			V\left(\mathbf{r}|\mathbf{R}_\text{L}\right)+\varepsilon &  x < -x_\text{L}',\\
			V\left(\mathbf{r}|\mathbf{R}_\text{M}\right)+\Delta & -x_\text{L}' < x < x_\text{R}' ,\\
			V\left(\mathbf{r}|\mathbf{R}_\text{R}\right)-\varepsilon & x > x_\text{R}',
		\end{cases}
	\end{equation}
where
\begin{equation}
	\begin{split}
		V(\mathbf{r}|\mathbf{R})=\frac{1}{2}m\omega_0^2 \left( \mathbf{r}-\mathbf{R}\right)^2,
	\end{split}
\end{equation}
$\mathbf{r}=(x,y)$ is the two dimensional vector in the plane of electron gas. $\mathbf{R}_\text{L}=(-x_0,0)$, $\mathbf{R}_\text{M}=(0,0)$ and $\mathbf{R}_\text{R}=(0,x_0)$ are the positions of the parabolic wells minima. $x'_\text{L}$ ($x'_\text{R}$) is the potential cut determined by locating the value of $x$ at which the potential values of the left (right) and middle dot are equal at $y=0$. $\Delta$ and $2\varepsilon$ are the middle-dot detuning and the relative detuning between the left and right dots, respectively. We denote the left, middle, and right dots as ``L'', ``M'', and ``R'', respectively. The effective mass $m^*$ is 0.067 electron mass in GaAs. $\omega_0$ is the confinement strength. $\mathbf{B}= B\mathbf{\hat{z}}$ is the perpendicular magnetic field and $\mathbf{S}$ is the total electron spin. 
Throughout this work, the magnetic field is set at $B=0.845$ T. The charge configuration of the ground state in the TQD is denoted as $(N_\text{L}N_\text{M}N_\text{R})$, where $N_j$ indicates number of electrons in dot $j$. Also, we denote a singlet (triplet) state formed by two antiparallel spins as $\vert S(N_\text{L}N_\text{M}N_\text{R})\rangle$ $\left[\vert T(N_\text{L}N_\text{M}N_\text{R})\rangle\right]$. For example, a singlet state formed by two electrons with each electron occupying the left and right dots is denoted as $\vert S(101)\rangle$.

We solve the Hamiltonian $H$ using the full CI technique \cite{Barnes.11}. In this work, we use the solutions to a single-particle Hamiltonian with a parabolic potential function in the presence of magnetic field, i.e. the Fock-Darwin (F-D) states, as the single-particle basis. For a given number of F-D states, we construct all possible two-electron Slater determinants and write down the explicit matrix form of the Hamiltonian $H$. We can then solve the problem by diagonalizing $H$ (see Sec. I in the Supplemental Material \cite{sm} for details). The numerical results of the system are obtained by keeping 10 orthonormalized F-D states in each dot, resulting in a total of 30 F-D states in a TQD device for the CI calculations. As suggested by the convergence of superexchange energy $J_\mathrm{se}$, this setup is sufficient to accurately simulate a two-electron system in the TQD device. 
Before we present the numerical results of full CI in Sec.~\ref{sec:result}, we first obtain an analytical result of $J_\text{se}$ based on the generalized form of Hubbard model \cite{Wang.11}, as shown in the following subsection, Sec. \ref{subsec:Jse}.

	
	\subsection{Analytical analysis of $J_\mathrm{se}$}\label{subsec:Jse}
	
		To understand the qualitative behavior of $J_\text{se}$ from full CI results, we derive the approximated form of the superexchange energy. The original Hamiltonian, Eq.~\eqref{eq:H}, can be written in the second quantization form (see Sec.~II in the Supplemental Material for details), resulting in a generic Hubbard model \cite{Wang.11}, which includes all possible Coulomb interactions terms. In addition, we have found that the low-lying physics of the superexchange energy from CI results can be well described using an effective Hamiltonian in which only the lowest orbital in each dot is considered. In this case, the generic Hubbard Hamiltonian can be simplified as follow
	\begin{equation}\label{eq:HubbardModel2ndQuan}
		\begin{split}
			H_\text{Hubbard} &= \sum_{j \sigma}\mu_{j \sigma} c^\dagger_{j \sigma} c_{j \sigma}+\sum_{j< k} \left(t_{j,k} c^\dagger_{j\sigma} c_{k\sigma}+\mathrm{H. c.}\right)
			\\
			&\quad+\sum_{j} U_{j} n_{j\downarrow} n_{j\uparrow} +\sum_{\sigma \sigma'}\sum_{j< k} U'_{j,k} n_{j\sigma} n_{k\sigma'}\\
			&\quad -\sum_{\sigma\neq\sigma'}\sum_{j< k}J^e_{j,k}c^\dagger_{j \sigma}c^\dagger_{k \sigma'}c_{k \sigma}c_{j \sigma}
			,
		\end{split}
	\end{equation}
	In Eq.~\eqref{eq:HubbardModel2ndQuan}, $j,k\in\{\text{L}, \text{M}, \text{R}\}$ refer to the orbitals in left (L), middle (M) and right (R) dots, respectively, while  $\sigma,\sigma'$ refer to the spins $\sigma,\sigma'\in\{\uparrow,\downarrow\}$. In the above generic Hubbard model, since only the lowest orbital in each dot is considered, the orbital index $\alpha$ is suppressed.
	$\mu_{j \sigma}$ is the chemical potential of an electron in dot $j$, $t_{j,k}$ is the tunneling energy between dot $j$ and $k$, $U_j$ is the on-site Coulomb interaction in dot $j$, $U'_{j,k}$ is the inter-dot Coulomb interaction between dot $j$ and $k$, $J^e_{j,k}$ is the Coulomb exchange interaction between dot $j$ and $k$. In Eq.~\eqref{eq:HubbardModel2ndQuan}, we have dropped other Coulomb-related terms which are negligible.
	The condition $j<k$ refers to the ordering from left to right. The detuning values are $2\varepsilon=\mu_\text{L} - \mu_\text{R}$ and $\Delta=\mu_2-(\mu_\text{L}+\mu_\text{R})/2$.

	For a two-electron system hosted in a TQD, $J_\mathrm{se}$ is defined as the energy splitting between the lowest singlet and triplet states in the $(N_\text{L}N_\text{M}N_\text{R})=(101)$ region. Following the experimental demonstrations of superexchange interaction \cite{JseDef}, we only consider $J_\text{se}$ in the detuning region where the ground charge state is (101). 
	We use the notation $\vert \eta \rangle'$ to denote an eigenstate whose main composition is $\vert \eta \rangle$, i.e.~$\langle \eta \vert \eta\rangle'\approx 1$. Also, the energy of state $\vert \eta \rangle$ is denoted as $E_{\vert \eta \rangle}$. With the notations defined above, $J_\mathrm{se}=E_{\vert T(101)\rangle'}-E_{\vert S(101)\rangle'}$. In the following, we introduce the notation $t \xi$ to denote the energy shifts derived from the perturbation theory, where $t=t_\text{L,M}=t_\text{M,R}$ is the nearest-neighbor tunneling and $\xi$ is the ratio of $t$ to the energy differences between states. Alternatively, the energy shifts are represented to be proportional to the admixture probabilities of excited states in the logical eigenstates, which we denote as $\phi$.
	
	The discussions on the analytical expressions of superexchange based on the generic Hubbard model,  denoted as $J_\text{se}^\text{Hubbard}$, are divided into two cases: (1) $\varepsilon\geq0$ while the values of $\left|\varepsilon\right|$ is small. In this case, $E_{\vert S(020)\rangle}>E_{\vert S(002)\rangle}>E_{\vert S(011)\rangle}>E_{\vert S(101)\rangle}$ \cite{EnergyOrder}, cf.~Fig.~\ref{fig:lvlDia}(a); (2) $\varepsilon\gg 0$. In this case, $E_{\vert S(020)\rangle}>E_{\vert S(011)\rangle}>E_{\vert S(002)\rangle}>E_{\vert S(101)\rangle}$, cf.~Fig.~\ref{fig:lvlDia}(b).  These two cases are discussed separately in Sec.~\ref{sec:JseAnalrPve} and Sec.~\ref{sec:JseAnalrDLarge}. 
	
	\subsubsection{$J_\mathrm{se}^\mathrm{Hubbard}$ for $\varepsilon\geq0$}\label{sec:JseAnalrPve}
	Based on the generic Hubbard model \cite{Wang.11,extendedHubbard}, the analytical expression of superexchange for $\varepsilon\geq0$ is (see Sec.~III A in the Supplemental Material \cite{sm} for derivations)
	
	\begin{equation}\label{eq:JseExp1Simpl}
		\begin{split}
			J_\text{se}^\text{Hubbard} \left(\Delta,\varepsilon\geq 0 \right) &\approx -2J^e_\text{L,R}+\left(t\xi_{t,\text{nn}}-2J^e_\text{L,M}\right)\times \xi_{t,\text{se}}^2
			\\
			&=-2J^e_\text{L,R}+J_\text{nn}\times \xi_\text{t,se}^2,
		\end{split}
	\end{equation}
	where
	\begin{subequations}\label{eq:JseExp1Simpl2}
		\begin{align}
			\begin{split}
				J_\text{nn}&=t\times \xi_{t,\text{nn}}-2J^e_\text{L,M},
			\end{split}
			\\
			\begin{split}
				\xi_{t,\text{nn}}&=\frac{ 2t}{E_{\vert S(200)\rangle}-E_{\vert S(110)\rangle}},
			\end{split}
			\\
			\begin{split}
				\xi_{t,\text{se}}&=\frac{t}{E_{\vert S(110)\rangle}-E_{\vert S(101)\rangle}},
			\end{split}
		\end{align}
	\end{subequations}
$J^e_\text{L,M}$ $\left(J^e_\text{L,R}\right)$ is the nearest-neighbor (next nearest-neighbor) Coulomb exchange energy between two neighboring (outmost) dots.

In Eq.~\eqref{eq:JseExp1Simpl}, the terms in the first round bracket in the first line denote the nearest-neighbor exchange coupling that occurs virtually in the excited energy states, cf.~the dashed blue double-headed arrow with the notation $\xi_{t,\text{nn}}$ in Fig.~\ref{fig:lvlDia}(a). In analogy to the nearest-neighbor exchange in the singly-occupied dots in a double-quantum-dot (DQD) device \cite{Li.12,Li.10,Burkard.99}, we denote those terms as $J_\text{nn}$, where ``nn'' indicates ``nearest-neighbor''. The term $\xi_{t,\text{se}}$ denotes, in addition to $J_\text{nn}$, the higher-order tunneling contribution to the superexchange $J_\text{se}^\text{Hubbard}$, cf.~the solid blue and red double-headed arrows with the notation $\xi_{t,\text{se}}$ in Fig.~\ref{fig:lvlDia}(a). Note that $J^e_\text{L,R}$ and $J_\text{nn}$ yield the energy unit while $\xi_{t,\text{nn}}$ and $\xi_{t,\text{se}}$ yield the unity unit. 

When $\varepsilon\geq0$, Eq.~\eqref{eq:JseExp1Simpl} indicates that $J_\text{se}^\text{Hubbard}$ results from the interplay between the virtual nearest-neighbor exchange term $J_\text{nn}$, the higher-order tunneling induced term $\xi_{t,\text{se}}$, and the long-distance Coulomb exchange term between two outmost dots $J^e_\text{L,R}$. 

\subsubsection{$J_\mathrm{se}^\mathrm{Hubbard}$ for $\varepsilon\gg 0$}\label{sec:JseAnalrDLarge}
Using the generic Hubbard model, the analytical expression of superexchange for $\varepsilon\gg0$ is (see Sec.~III B in the Supplemental Material \cite{sm} for derivations)
\begin{align}\label{eq:JseExp2}
\begin{split}
	&J_\text{se}^\text{Hubbard}(\Delta,\varepsilon\gg0)
	\\&=-2J^e_\text{L,R}+t \left(\xi_{tt,\text{se}}+\xi_{t,\text{LM}}\right)
	\\
	&\propto -2J^e_\text{L,R}+t \left(\phi_{\vert S(002)\rangle}
	+\delta \phi_{\vert (011)\rangle}\right),
\end{split}
\end{align}
where
\begin{equation}
		\begin{split}
			\phi_{\vert S(N_\text{L}N_\text{M}N_\text{R})\rangle}&=\left| \langle S(N_\text{L}N_\text{M}N_\text{R}) \vert S(101)\rangle'\right|^2,
		\\
			\phi_{\vert T(N_\text{L}N_\text{M}N_\text{R})\rangle}&=\left| \langle T(N_\text{L}N_\text{M}N_\text{R}) \vert T(101)\rangle'\right|^2,
	\\
		\delta \phi_{\vert(011)\rangle}&=\left( \phi_{\vert S(011)\rangle}-\phi_{\vert T(011)\rangle}\right),
	\end{split}
\end{equation}
while
\begin{equation}
	\begin{split}
		\xi_{tt,\text{se}}&=\left(\frac{{t}}{\Delta E_{\vert S(011)\rangle,\vert S(002)\rangle} }\right)^2
		\frac{t}{\Delta E_{\vert S(002)\rangle,\vert S(101)\rangle}},
		\\
		\xi_{t,\text{LM}}&=\frac{t}{\Delta E_{\vert S(011)\rangle,\vert S(101)\rangle}}
		-\frac{t}{\Delta E_{\vert T(011)\rangle,\vert T(101)\rangle}},
	\end{split}
\end{equation}
$\Delta E_{\vert \eta_1 \rangle,\vert \eta_2 \rangle}=E_{\vert \eta_1 \rangle}-E_{\vert \eta_2 \rangle}$.
In Eq.~\eqref{eq:JseExp2}, the term $t \times \xi_{tt,\text{se}}$ $\left(t \times \xi_{t,\text{LM}}\right)$ denotes a four-tunneling process (two-tunneling processes), cf.~the solid blue double-headed arrow (the dashed blue and red double-headed arrows) in Fig.~\ref{fig:lvlDia}(b). Note that $J_\text{se}^\text{Hubbard}$ yields the energy unit while $\xi_{tt,\text{se}}$ and $\xi_{t,\text{LM}}$ yield the unity unit. The tunneling processes in Fig.~\ref{fig:lvlDia}(b) indicate that $J_\text{se}^\text{Hubbard}$ can be made proportional to the admixture probabilities of the excited states in the logical eigenstates, $\phi_{\vert \eta \rangle}$, cf.~the third line in Eq.~\eqref{eq:JseExp2}. Similar to the physical mechanism of $\xi$, $\phi_{\vert S(002)\rangle}$ $\left[\phi_{\vert(011)\rangle}\right]$ arises from a four-tunneling process (two-tunneling processes). 

\begin{figure*}[t]
	\includegraphics[width=\linewidth]{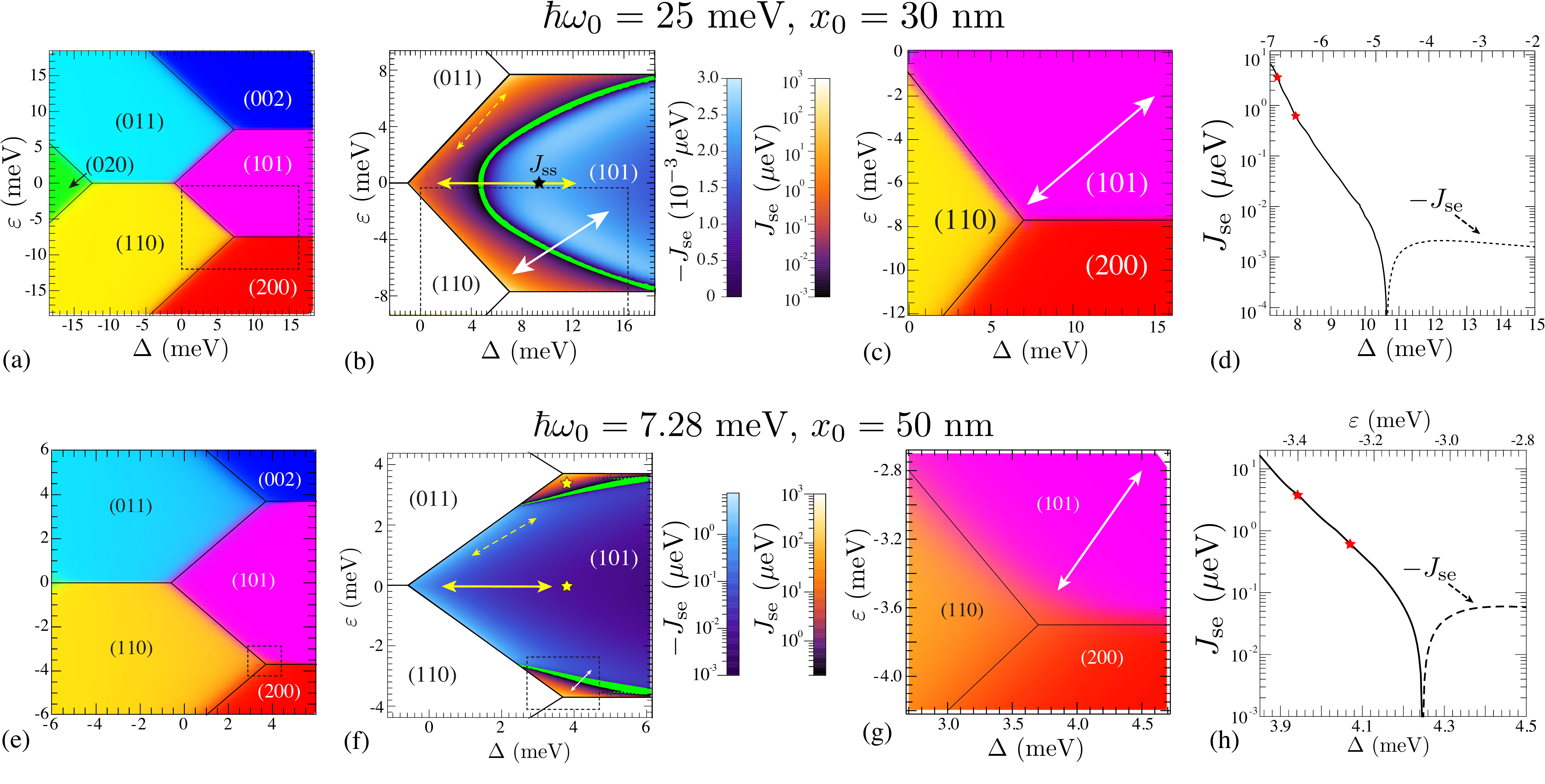}
	\caption{Stability diagrams and $J_\text{se}$ for (a)-(d) $\hbar \omega_0 = 25$ meV, $x_0 = 30$ nm and (e)-(h) $\hbar \omega_0 = 7.28$ meV, $x_0 = 50$ nm. (a), (e) Stability diagrams with the RGB color code defined as $\left(128\times N_\text{L},128\times N_\text{M},128\times N_\text{R}\right)$. (b), (f) $J_\text{se}$ v.s.~$\Delta$ and $\varepsilon$ in the (101) region. The green region marks the transition from $J_\text{se}>0$ to $J_\text{se}<0$. $J_\text{ss}$ denotes the sweet spot with respect to $\Delta$ and $\varepsilon$, i.e.~$\partial J_\text{se}/\partial \Delta=\partial J_\text{se}/\partial \varepsilon=0$. (c) and (g) show the zoom in view of (a) and (e), respectively, as indicated by the dashed line boxes. (d), (h) $J_\text{se}$ v.s.~$\Delta$ and $\varepsilon$. The parings of $\Delta$ and $\varepsilon$ are given by the white double-headed arrows in (c) and (g). Two values of $J_\text{se}$ demonstrated in Ref.~\onlinecite{Baart.17} are presented as red stars.}
	\label{fig:staDiaandJ}
\end{figure*}

In this work, the analytical expressions of superexchange, denoted as $J_\text{se}^\text{Hubbard}$ in Eqs.~\eqref{eq:JseExp1Simpl} and  \eqref{eq:JseExp2}, are adopted to provide qualitative descriptions for the behavior of superexchange energy. Note the different notations between $J_\text{se}^\text{Hubbard}$ and $J_\text{se}$ in the following sections: the former describes the analytical behavior of superexchange energy in terms of Hubbard parameters while the latter is the exact numerical results using full CI calculations.

\section{Results}\label{sec:result}
In this section, we discuss the values for $J_\text{se}$ in different parameter regimes. For illustration purposes, throughout this paper, we plot $J_\text{se}$ with positive and negative values as solid and dashed lines, respectively. It should be noted that, throughout this paper, the numerical values of $J_\text{se}$ are obtained directly from full CI results without mapping to the generic Hubbard model, i.e.~taking the energy difference between the lowest singlet and triplet states. 
Other quantities, e.g.~the tunneling terms $\xi$ and admixture probabilities $\phi$ in Eqs.~\eqref{eq:JseExp1Simpl} and \eqref{eq:JseExp2}, are obtained by mapping to the generic Hubbard model, which allows us to comprehend the behaviors of $J_\text{se}$ from the perspective of the analytical expressions of $J_\text{se}^\text{Hubbard}$ (see Secs.~IV and V in the Supplemental Material for the details on evaluating the values of $\xi$ and $\phi$, respectivaly). 
It should be noted that the numerical values shown in the following figures, including $J_\text{se}$, $\xi$, and $\phi$, are extracted directly from the CI results. The only approximation made is the qualitative descriptions of the behavior of $J_\text{se}$ based on the changes in the values of $\xi$ and $\phi$ and the expressions of $J_\text{se}^\text{Hubbard}$ in Eqs.~\eqref{eq:JseExp1Simpl} and \eqref{eq:JseExp2}.
The values of $J_\text{se}$ are obtained in the same way as the CI methods in Ref.~\onlinecite{Deng.20}. On the other hand, it is in contrast to Ref.~\onlinecite{Rancic.17} in which $J_\text{se}$ is evaluated from Hubbard model.


\subsection{$J_\mathrm{se}$ v.s.~$\Delta$ and $\varepsilon$}\label{sec:JseVSlrmD}
Figure \ref{fig:staDiaandJ} shows the results for (a)-(d) $\hbar \omega_0=25$ meV, $x_0= 30$ nm and (e)-(h) $\hbar \omega_0=7.28$ meV, $x_0= 50$ nm. Note that, in Figs.~\ref{fig:staDiaandJ}(b), (d), (f), and (h), the values of $J_\text{se}$ are only provided in the detuning regimes where the ground charge states are (101), cf.~the stability diagrams in Figs.~\ref{fig:staDiaandJ}(a) and (e).

To compare our findings with experimental results, $J_\text{se}$ is extracted along the double-headed white arrows as shown in Figs.~\ref{fig:staDiaandJ}(c) and (g), with the corresponding values of $J_\text{se}$ provided in Figs.~\ref{fig:staDiaandJ}(d) and (h), respectively. Figs.~\ref{fig:staDiaandJ}(d) and (h) show that, within the part that gives $J_\text{se}>0$, $J_\text{se}$ yields a monotonic decrease along the white arrows in Fig.~\ref{fig:staDiaandJ}(c) and (g). In addition, the magnitude of $J_\text{se}$ includes the values of $J_\text{se}$ demonstrated in Ref.~\onlinecite{Baart.17}. In Ref.~\onlinecite{Baart.17}, at the detuning value that is close to the triple-transition-point of the $(200)$, $(110)$ and $(101)$ regions, the system exhibits a strong $J_\text{se}$ with a magnitude of $\sim3.75$ $\mu$eV (900 MHz). On the other hand, at the detuning value that is away from the triple-transition-point, the system exhibits a weak $J_\text{se}$ with a magnitude of $\sim0.625$ $\mu$eV (150 MHz). In the region where $J_\text{se}>0$, as shown in Fig.~\ref{fig:staDiaandJ}(d) and (h), the values of $J_\text{se}$ that are positive and decrease monotonically as a function of $\varepsilon$ agree with results presented in Ref.~\onlinecite{Baart.17}, cf.~the red star symbols.

In addition, Fig.~\ref{fig:staDiaandJ}(b) and (f) show that the values of $J_\text{se}$ exhibit a non-trivial behavior when a larger detuning region is considered. In particular, the behavior of $J_\text{se}$ can be divided into three cases, as enumerated in the following paragraphs.

First, we consider the values of $J_\text{se}$ at $\varepsilon=0$. For a large $\hbar \omega_0$ and smaller $x_0$, the values of $J_\text{se}$ experience a sign-switching event as a function of $\Delta$, cf.~the solid yellow double-headed arrow in Fig.~\ref{fig:staDiaandJ}(b). On the other hand, for a small $\hbar \omega_0$ and larger $x_0$, the values of $J_\text{se}$ are negative in all values of $\Delta$, cf.~the solid yellow double-headed arrow in Fig.~\ref{fig:staDiaandJ}(f). The qualitative analyses of this behavior of $J_\text{se}$ will be provided in Sec.~\ref{sec:JseDeltaLR0}.

\begin{figure*}[tbp]
	\includegraphics[width=\linewidth]{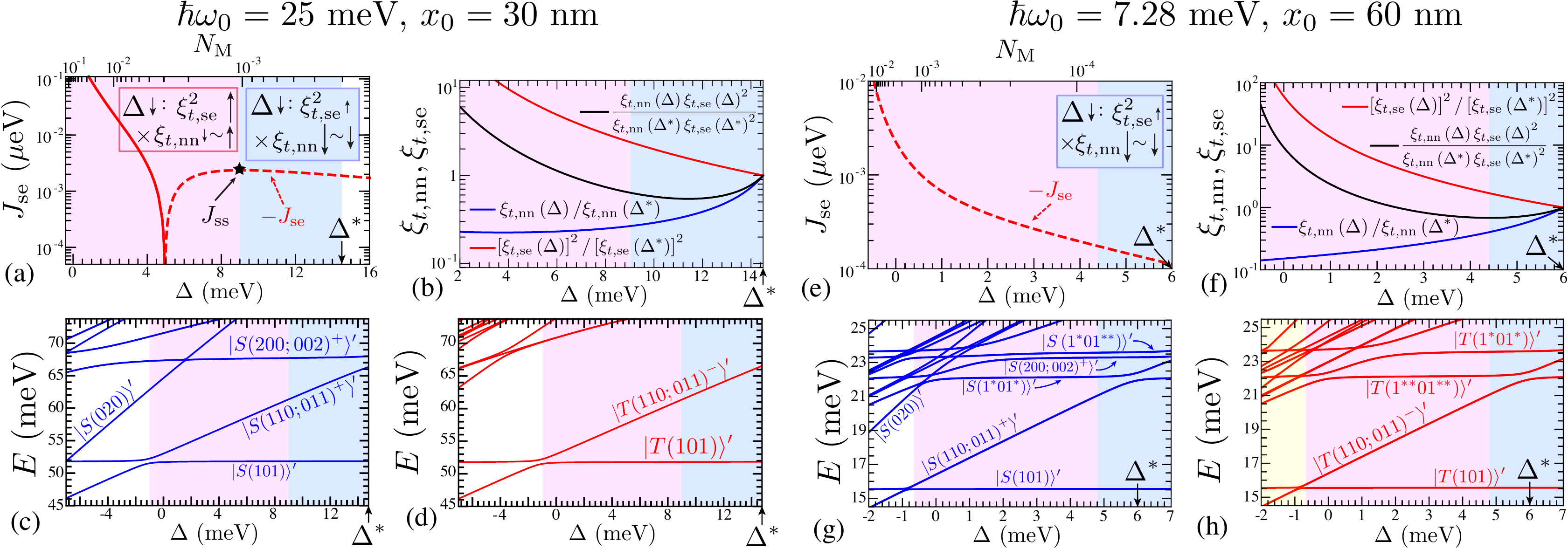}
	\caption{Results of $J_\text{se}$, eigenvalues of singlet and triplet states, and $\xi_t$ v.s.~$\Delta$. The results are obtained for $\varepsilon=0$ with dot parameters (a)-(d) $\hbar\omega_0=25$ meV, $x_0=30$ nm and (e)-(h) $\hbar\omega_0=7.28$ meV, $x_0=60$ nm. (a), (e) Superexchange energy $J_\text{se}$ v.s.~$\Delta$. Positive and negative $J_\text{se}$ are plotted as solid and dashed lines, respectively. $J_\text{ss}$ denotes the sweet spot. The bottom axes label the $\Delta$ values with the corresponding values of $N_\text{M}$ are labeled at the top axes. The arrows in the insets denote the changes of $\xi_{t,\text{se}}$ and $\xi_{t,\text{nn}}$ with respect to $\Delta$, while the lengths of the arrows indicate the magnitude of the changes. (b), (f) $\xi_{t,l}$ v.s.~$\Delta$ with $l \in \{\text{nn},\text{se}\}$. The values are presented as ratios to those at $\Delta=\Delta^*$, i.e. $\xi_{t,{l}}\left(\Delta\right)=\xi_{t,{l}}\left(\Delta\right)/\xi_{t,{l}}\left(\Delta^*\right)$. Lowest eigenvalues of ((c), (g)) singlets and ((d), (h)) triplets v.s.~$\Delta$. At $\varepsilon=0$, since $E_{\vert S(110)\rangle'}=E_{\vert S(011)\rangle'}$ $\left[E_{\vert S(200)\rangle'}=E_{\vert S(002)\rangle'}\right]$, they form a hybridized state $\vert S(110;011)^+\rangle'=(\vert S(011)\rangle'+\vert S(011)\rangle')/\sqrt{2}$ $\left[\vert S(200;002)^+\rangle'=(\vert S(200)\rangle'+\vert S(002)\rangle')/\sqrt{2}\right]$. Also, since $E_{\vert T(110)\rangle'}=E_{\vert T(011)\rangle'}$, they form a hybridized state $\vert T(110;011)^-\rangle'=(\vert T(011)\rangle'-\vert T(011)\rangle')/\sqrt{2}$. The number of asterisks (*) indicate the degrees of excitation from the ground configurations. 
}
	\label{fig:JvsDeltaTwoCases}
\end{figure*}

Second, we consider the values of $J_\text{se}$ for $\varepsilon>0$. The detuning region where $\varepsilon>0$ is defined when the ordering of the energies of singlet states is $E_{\vert S(020)\rangle}>E_{\vert S(002)\rangle}>E_{\vert S(011)\rangle}>E_{\vert S(101)\rangle}$, cf.~Sec.~\ref{sec:JseAnalrPve} and Fig.~\ref{fig:lvlDia}(a). For a large $\hbar \omega_0$ and smaller $x_0$, under a fixed value of $N_\text{M}$, the magnitude of an originally positive $J_\text{se}$ increases for a larger $\varepsilon$, cf.~dashed double-headed arrows in Fig.~\ref{fig:staDiaandJ}(b). On the other hand, for a small $\hbar \omega_0$ and larger $x_0$, under a fixed value of $N_\text{M}$, the magnitude of an originally negative $J_\text{se}$ decreases for a larger $\varepsilon$, cf.~dashed double-headed arrows in Fig.~\ref{fig:staDiaandJ}(f). Note that the changes of the magnitude of $J_\text{se}$ are less obvious in the density plots in Figs.~\ref{fig:staDiaandJ}(b) and (f). More evident comparisons on the values of $J_\text{se}$ for different $\varepsilon$ will be provided in Sec.~\ref{subsubsec:JseS002lS011in101}, where the qualitative analyses of  $J_\text{se}$ are also included. 

Third, we consider the values of $J_\text{se}$ when $\varepsilon\gg0$. The detuning region where $\varepsilon\gg0$ is defined when the ordering of the energies of singlet states is $E_{\vert S(020)\rangle}>E_{\vert S(011)\rangle}>E_{\vert S(002)\rangle}>E_{\vert S(101)\rangle}$, cf.~Sec.~\ref{sec:JseAnalrDLarge} and Fig.~\ref{fig:lvlDia}(b). For a small $\hbar\omega_0$ and larger $x_0$, a negative $J_\text{se}$ switches to a positive value when $\varepsilon$ is largely increased, cf.~yellow star symbols in Fig.~\ref{fig:staDiaandJ}(f).  The qualitative analyses of this behavior of $J_\text{se}$ will be provided in Sec.~\ref{subsubsec:JswitchDLRLarge}.

\subsection{$J_\mathrm{se}$ v.s.~$\Delta$ at $\varepsilon=0$}\label{sec:JseDeltaLR0}.

\subsubsection{$J_\mathrm{se}$ switches sign in $\mathrm{(101)}$ region}\label{subsubsec:hbLx0m}
In this subsection, we discuss the sign-switching event observed for $J_\mathrm{se}$ at $\varepsilon=0$, cf.~the yellow solid double-headed arrow in Fig.~\ref{fig:staDiaandJ}(b). The sign-switching event of $J_\text{se}$ is found for a TQD device with a large $\hbar \omega_0$ and smaller $x_0$, cf.~Fig.~\ref{fig:staDiaandJ}(b) and Fig.~\ref{fig:staDiaandJ}(f).
To understand the physical mechanism giving rise to the sign-switching of $J_\text{se}$, the explicit values of $J_\text{se}$ and other relevant quantities at $\varepsilon=0$ are presented in Fig.~\ref{fig:JvsDeltaTwoCases}.
The following discussion on the numerical values of $J_\text{se}$ is made in reference to the results presented in the left half of Fig.~\ref{fig:JvsDeltaTwoCases}, i.e.~Fig.~\ref{fig:JvsDeltaTwoCases}(a)-(d). 

It is helpful to start the discussion at the large $\Delta$ region ($\Delta>14.5$ meV), which shows that $J_\text{se}<0$, cf.~Fig.~\ref{fig:JvsDeltaTwoCases}(a). This can be understood using Eq.~\eqref{eq:JseExp1Simpl}, which gives $J^\text{Hubbard}_\text{se}\approx -2J^e_\text{L,R}$ as $\xi_{t,\text{se}} \approx 0$ for $E_{\vert S(110)\rangle}-E_{\vert S(101)\rangle}\gg 0$. 
This is in contrast to a two-electron system in a DQD in which energy shifts on the singly-occupied states are present for the singlet state while forbidden for the triplet state due to Coulomb blockade. This is also in contrast to the superexchange energy evaluated using Hubbard model \cite{Rancic.17}, which neglects the Coulomb exchange energy $J^e_\text{L,R}$.

Next, we focus at the region where $\Delta$ is moderately large, i.e.~9 meV $<\Delta < $ 14.5 meV (the light blue background region). Fig.~\ref{fig:JvsDeltaTwoCases}(a) shows that $\left|J_\text{se}\right|$ increases when $\Delta$ decreases. This can be attributed to the decrease of $\xi_{t,\text{nn}} \xi_{t,\text{se}}^2$ for a decreasing $\Delta$, cf.~Eq.~\eqref{eq:JseExp1Simpl}, conforming with the decrease of black line in the light blue region in Fig.~\ref{fig:JvsDeltaTwoCases}(b). 

Lastly, we inspect the small $\Delta$ region, i.e.~0 meV $<\Delta<9$ meV), as indicated by the magenta background. Fig.~\ref{fig:JvsDeltaTwoCases}(a) shows that, starting from $\Delta = 9$ meV, when $\Delta$ decreases, an initially negative $J_\text{se}$ increases until it switches to positive $J_\text{se}$, which continues to increase for smaller $\Delta$. Fig.~\ref{fig:JvsDeltaTwoCases}(b) (magenta background) shows that the increase of $J_\text{se}$ with decreasing $\Delta$ can be attributed to the increase of $\xi_{t,\text{nn}}  \xi_{t,\text{se}}^2$ (black line). 
In that detuning region, Fig.~\ref{fig:JvsDeltaTwoCases}(b) shows that the increase of $\xi_{t,\text{nn}}  \xi_{t,\text{se}}^2$ (black line) is dominated by the substantial increase of $\xi_{t,\text{se}}^2$ (blue line).

To conclude this subsection, we have shown, for a large $\hbar \omega_0$ and smaller $x_0$, that $J_\text{se}$ is negative at large $\Delta$ and switches to a positive value at small $\Delta$. We have found that the sign-switching event is due to the enhancement of the higher-order tunneling induced term $\xi_{t,\text{se}}$, which originates from the tunneling between the lowest two singlet states and the lowest two triplet states, cf.~Fig.~\ref{fig:lvlDia}(a). In addition, the non-monotonic behavior of overall energy shift $\left(t \xi_{t,\text{nn}} \xi_{t,\text{se}}^2\right)$ leads to a non-monotonous $J_\text{se}$, giving rise to a sweet spot with respect to $\Delta$ and $\varepsilon$, cf.~the black star symbol in Fig.~\ref{fig:JvsDeltaTwoCases}(a).

Before moving on, it is worth commenting on the existence of the zero-crossing and non-monotonic behavior of $J_\text{se}$ in terms of the choice of confinement model. First, as mentioned previously, the negative $J_\text{se}$ in a TQD arises at large $\Delta$ regardless of the choice of confinement model. If another potential model, such as a quartic model \cite{White.18,Culcer.09}, is chosen, the interdot tunneling would be enhanced, such that a larger range of $\Delta$ exhibits $J_\text{se} > 0$. This is because the zero-crossing occurs when the magnitude of the overall energy shift ($t \xi_{t,\text{se}}^2\xi_{t,\text{nn}}$), which is proportional to the inter-dot tunneling, exceeds the magnitude of Coulomb exchange energy ($\left|-2J^e_\text{L,R}\right|$). Second, analyses above show that the non-monotonous $J_\text{se}$ arises from the interplay between different tunneling terms $\xi$, whose denominators are the energy differences between states, cf.~Eq.~\eqref{eq:JseExp1Simpl2}. Therefore, the non-monotonous of $J_\text{se}$ can be alternatively interpreted as the interplay between different energy differences in terms of $\Delta$, i.e.~the increase of $ E_{\vert S(200)\rangle}-E_{\vert S(110)\rangle}$ and the decrease of $ E_{\vert S(110)\rangle}-E_{\vert S(101)\rangle}$ for a decreasing $\Delta$. Such changes in the energy differences apply to any confinement model. The above discussions argue that the non-monotonous and zero-crossing of $J_\text{se}$ could be observed regardless of the choice of confinement model. However, confirming the claim quantitatively is currently out of the scope of this work.


\subsubsection{$J_\mathrm{se}<0$ in $\mathrm{(101)}$ region}\label{subsubsec:hbmx0L}
In this subsection, we provide analyses to understand the negative $J_\text{se}$ at $\varepsilon=0$ as indicated by the yellow solid double-headed arrow in Fig.~\ref{fig:staDiaandJ}(f). For a small $\hbar \omega_0$ and larger $x_0$, negative $J_\text{se}$ is observed in the whole range of $\Delta$ in the (101) region, cf.~Fig.~\ref{fig:staDiaandJ}(b) and Fig.~\ref{fig:staDiaandJ}(f). The corresponding explicit values of $J_\text{se}$ and other relevant quantities are provided in the right half of Fig.~\ref{fig:JvsDeltaTwoCases} \cite{x060}, i.e.~Figs.~\ref{fig:JvsDeltaTwoCases}(e)-(h). In particular, Fig.~\ref{fig:JvsDeltaTwoCases}(e) shows, in the (101) region, that $J_\text{se}<0$ in the whole range of $\Delta$. 

The qualitative behavior of the negative $J_\text{se}$ as a function of $\Delta$ and its description based on the magnitude of $\xi_{t,{\text{nn}}}\xi_{t,{\text{se}}}^2$ follow the discussions in the previous subsection, Sec.~\ref{subsubsec:hbLx0m}. However, different from the values of $J_\text{se}$ for a large $\hbar\omega_0$ and smaller $x_0$ [Fig.~\ref{fig:JvsDeltaTwoCases}(a)], Fig.~\ref{fig:JvsDeltaTwoCases}(e) shows that $J_\text{se}$ is always negative and exhibits a monotonic behavior. This can be attributed to the smaller nearest-neighbor tunneling value $t$ observed for the system. Fitting the full CI results to the generic Hubbard model shows that $t=255$ $\mu$eV for the previous system ($\hbar \omega_0=25$ meV and $x_0=30$ nm) while $t=33$ $\mu$eV for the current system of interest ($\hbar \omega_0=7.28$ meV and $x_0=60$ nm). As discussed in the previous subsection, the increase of $\xi_{t,\text{se}}^2$ is essential in switching an originally negative $J_\text{se}$ to a positive one, cf.~Eq.~\ref{eq:JseExp1Simpl}. As $\xi \propto t$, a relatively small $t$ leads to a smaller $\xi_{t,\text{se}}^2$, resulting in $J_\text{se}$ remaining negative in the whole range of $\Delta$ in the (101) region, cf.~Fig.~\ref{fig:JvsDeltaTwoCases}(e).

\begin{figure}[t]
	\includegraphics[width=\linewidth]{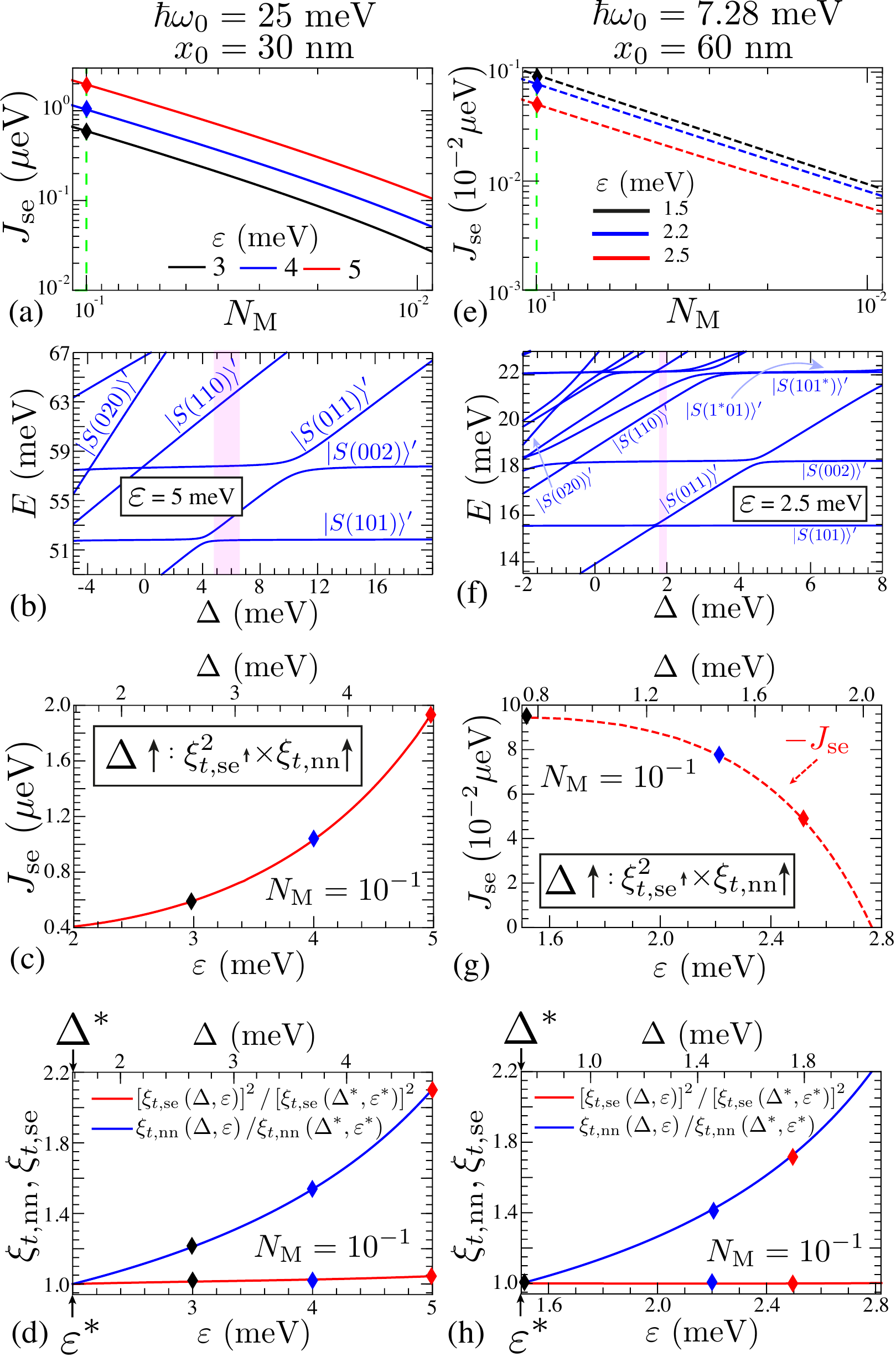}
	\caption{(a), (e) $J_\text{se}$ v.s.~$N_\text{M}$ for $\varepsilon>0$. The values of $N_\text{M}$ should be perceived as inversely proportional to $\Delta$. The values of $J_\text{se}$ at $N_\text{M}=10^{-1}$ are indicated as diamond symbols. (b), (f) Eigenvalues of the singlet states v.s.~$\Delta$ at a particular value of $\varepsilon$. (c), (g) $J_\text{se}$ v.s.~$\varepsilon$ for $N_\text{M}=10^{-1}$. The arrows in the insets indicate the changes of $\xi_{t,\text{se(L,M)}}$ and $\xi_{t,\text{nn}}$ with respect to $\Delta$, while the lengths of the arrows denote the magnitude of the changes. are indicated as diamond symbols. (d), (h) $\xi_{t,l}$ v.s.~$\varepsilon$ with $l\in\{\text{se},\text{nn}\}$. The values of $\xi_{t,l}$ are plotted as ratios to the values at $\varepsilon=\varepsilon^*$ and $\Delta=\Delta^*$, i.e. $\xi_{t,l}\left(\Delta,\varepsilon\right)/\xi_{t,l}\left(\Delta^*,\varepsilon^*\right)$. The top axes in (c, d, g, h) indicate the corresponding $\Delta$ for maintaining $N_\text{M}=10^{-1}$. The values of $J_\text{se}$ and $\xi_{t,l}$ for the values of $\varepsilon$ in (a) and (b) are denoted as diamond symbols.}
	\label{fig:JandStProbCompare2}
\end{figure}

To conclude this subsection, we have shown, for a small $\hbar \omega_0$ and larger $x_0$, that $J_\text{se}$ is negative in all ranges of $\Delta$ in the (101) region. We further show that $J_\text{se}$ maintains a negative value due to the smaller value of nearest-neighbor tunneling $t$.

In addition, a comparison between the behaviors of $J_\text{se}$ in Fig.~\ref{fig:JvsDeltaTwoCases}(a) and  Fig.~\ref{fig:JvsDeltaTwoCases}(e) suggests that a TQD device with a smaller $\hbar \omega_0$ and larger $x_0$ has a higher tendency to exhibit a negative $J_\text{se}$ or vice verse. Extensive CI calculations of $J_\text{se}$ for various values of $\hbar \omega_0$ (not shown here) conform with the claim above.


\subsection{$J_\mathrm{se}$ v.s. $\Delta$ for $\varepsilon>0$}\label{subsubsec:JseS002lS011in101}

In this section, we provide analyses on the values of $J_\text{se}$ for $\varepsilon>0$. In particular, we would like to understand the physical mechanism that gives rise to the changes of $J_\text{se}$ when the values of $\varepsilon$ changes, cf.~yellow dashed double-headed arrows in Figs.~\ref{fig:staDiaandJ}(b) and (f). The values of $J_\text{se}$ in Figs.~\ref{fig:staDiaandJ}(b) and (f) are extracted and provided in detail in Fig.~\ref{fig:JandStProbCompare2} \cite{x060}. The condition, $\varepsilon>0$, denotes the detuning regime where the order of the energies of singlet states is $E_{\vert S (002)\rangle}>E_{\vert S (011)\rangle}>E_{\vert S (101)\rangle}$, cf.~Figs.~\ref{fig:JandStProbCompare2}(b) and (f). 
In Figs.~\ref{fig:JandStProbCompare2}(a) and (e), we choose to plot the values of $J_\text{se}$ as a function of $N_\text{M}$ instead of $\Delta$ to compare different values of $\varepsilon$. This is due to the values of $J_\text{se}$ being determined by the hybridization between the logical states and higher-lying states, which is related to the electron densities in dot M. 


For a large $\hbar \omega_0$ and smaller $x_0$, Figs.~\ref{fig:JandStProbCompare2}(a) and (c) show that $J_\text{se}$ is positive and increases as a function of $\varepsilon$. The positive $J_\text{se}$ indicates, in those detuning regimes, that $J_\text{nn} \times  \xi_{t,\text{se}}^2>2J^e_\text{L,R}$, cf.~Eq.~\eqref{eq:JseExp1Simpl}. Therefore, the increase of $J_\text{se}$ can be attributed to the increase of $\xi_{t,\text{nn}}\xi_{t,\text{se}}^2$ as a function of $\varepsilon$, as shown in  Fig.~\ref{fig:JandStProbCompare2}(d).  On the other hand, for a small $\hbar \omega_0$ and larger $x_0$, Figs.~\ref{fig:JandStProbCompare2}(e) and (g) show that $J_\text{se}$ is negative and its absolute value decreases as a function of $\varepsilon$. The negative $J_\text{se}$ indicates, in those detuning regimes, that $J_\text{nn} \xi_{t,\text{se}}^2<2J^e_\text{L,R}$, cf.~Eq.~\eqref{eq:JseExp1Simpl}. Therefore, the decrease of the absolute value of $J_\text{se}$ can be attributed to the increase of $\xi_{t,\text{nn}} \xi_{t,\text{se}}^2$ as a function of $\varepsilon$, as shown in  Fig.~\ref{fig:JandStProbCompare2}(h). Since $J^\text{Hubbard}_\text{se}\approx-2J^e_\text{L,R} +\left(t' \xi_{t,\text{nn}}-2J^e_\text{L,M}\right)\xi_{t,\text{se}}^2$ (Eq.~\eqref{eq:JseExp1Simpl}), the increase of $\xi_{t,\text{nn}}\xi_{t,\text{se}}^2$ compensates the negative value of $-2J^e_\text{L,R}$, resulting in a smaller magnitude of superexchange, as shown in Fig.~\ref{fig:JandStProbCompare2}(g).

To conclude this section, we have shown, when the value of $\varepsilon$ increases, that the magnitude of $J_\text{se}$ increases (decreases) for an originally positive (negative) $J_\text{se}$. We further demonstrate, for an increasing $\varepsilon$, that the changes of the magnitude of $J_\text{se}$ can be attributed to the increase of the tunneling induced term $\xi_{t,\text{nn}}\xi_{t,\text{se}}^2$.

\subsection{$J_\mathrm{se}$ v.s. $\Delta$ at $\varepsilon\gg0$}\label{subsubsec:JswitchDLRLarge}

\begin{figure}[t]
	\includegraphics[width=\linewidth]{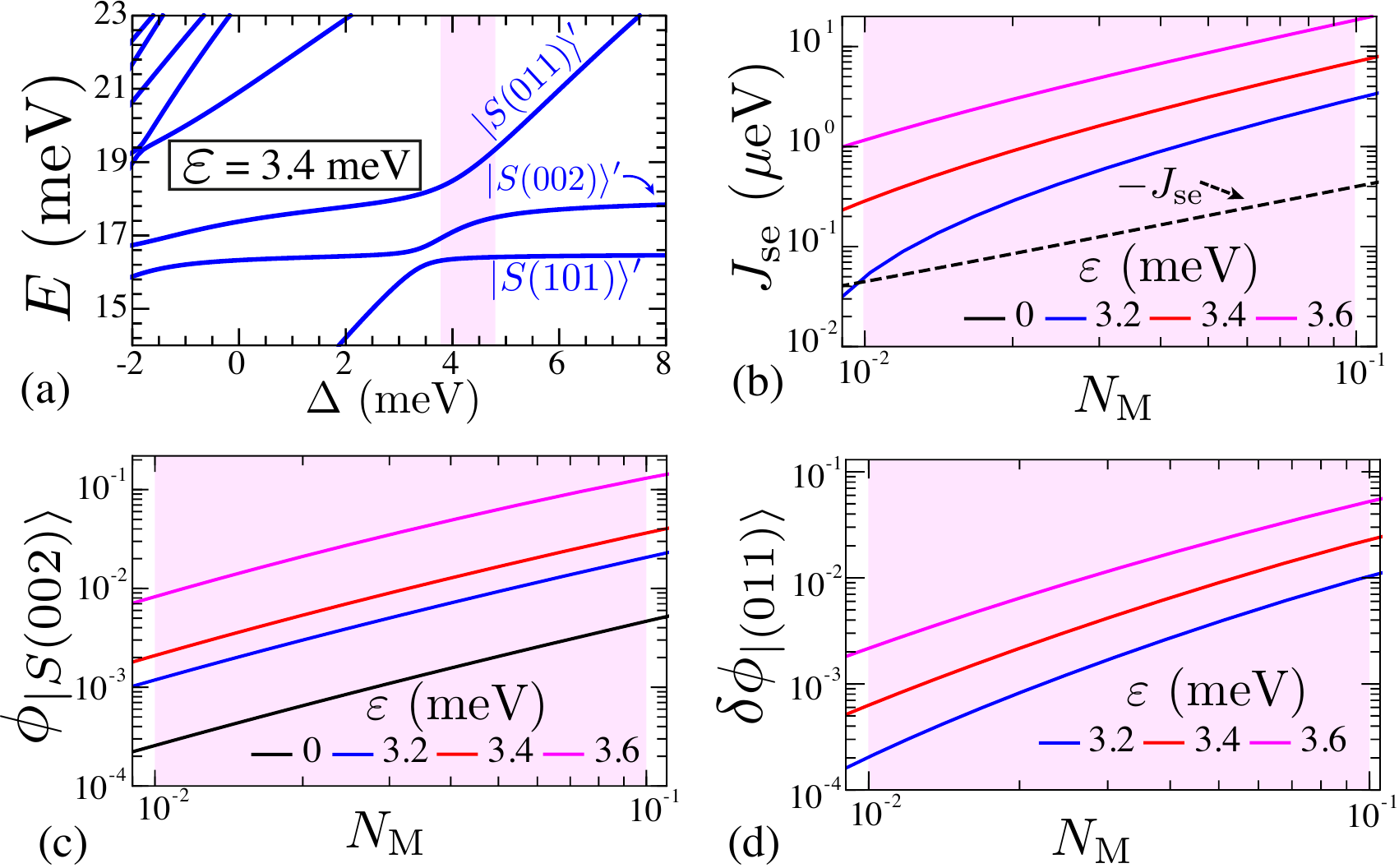}
	\caption{(a) Eigenvalues of singlet states v.s.~$\Delta$ for $\varepsilon=3.4$ meV. (b) $J_\text{se}$ v.s.~$N_\text{M}$ for $\varepsilon=0$ meV (black line), $3.2$ meV (blue line), $3.4$ meV (red line) and $3.6$ meV (magenta line). Solid and dashed lines indicate $J_\text{se}>0$ and $J_\text{se}<0$, respectively. (c) The composition of $\vert S\left(002\right)\rangle$ in $\vert S\left(101\right)\rangle'$, $\Phi_{\vert S(002)\rangle}$, v.s.~$N_\text{M}$ for the values of $\varepsilon$ in (b). Note that, since $E_{\vert S(200)\rangle}=E_{\vert S(002)\rangle}$ at $\varepsilon=0$, the black line refers to $\left|\langle S(200;002)^+ \vert S(101)\rangle'\right|^2$. (d) $\delta \phi_{\vert 011\rangle}$ v.s.~$N_\text{M}$ for the values of $\varepsilon$ in (b), where $\delta \phi_{\vert 011\rangle}=\phi_{\vert S(011)\rangle}-\phi_{\vert T(011)\rangle}$. $N_\text{M}$ should be perceived to be inversely proportional to $\Delta$. The values of $\delta \phi_{\vert 011\rangle}$ for $\varepsilon=0$ are not shown since $\delta \phi_{\vert 011\rangle}\ll 10^{-4}$. The magenta background indicated the (101) region where $10^{-2}<N_\text{M}<10^{-1}$. The dot parameters are: $\hbar \omega_0 = 7.28$ meV, $x_0=50$ nm.}
	\label{fig:JDeltaLRLarge}
\end{figure}

In this section, we provide analytical descriptions on the sign-switching of $J_\text{se}$ observed between $\varepsilon=0$ and $\varepsilon\gg0$, cf.~two yellow star symbols in Fig.~\ref{fig:staDiaandJ}(f). The explicit values of $J_\text{se}$ and other relevant quantities are presented in Fig.~\ref{fig:JDeltaLRLarge}.

In Fig.~\ref{fig:JDeltaLRLarge}(b), $\varepsilon=3.2\sim3.6$ meV corresponds to the case in which $\varepsilon\gg0$, as discussed in Sec.~\ref{sec:JseAnalrDLarge}. In particular, the condition $\varepsilon\gg0$ denotes the detuning regime where the order of the energies of singlet states is $E_{\vert S(011)\rangle}>E_{\vert S(002)\rangle}>E_{\vert S(101)\rangle}$, cf.~Fig.~\ref{fig:JDeltaLRLarge}(a). Fig.~\ref{fig:JDeltaLRLarge}(b) shows that a negative $J_\text{se}$ at $\varepsilon=0$ switches a positive value when $\varepsilon\gg0$. The sign-switching of $J_\text{se}$ from a negative to a positive value can be attributed to the larger values the admixture probabilities, $\phi_{\vert S(002)\rangle}$ and $\delta \phi_{\vert(011)\rangle}$,  cf.~Eq.~\eqref{eq:JseExp2}. The claim above conforms with the values shown in Figs.~\ref{fig:JDeltaLRLarge}(c) and (d), which shows more than one order of increase for $\phi_{\vert S(002)\rangle}$ and a large magnitude enhancement for $\delta \phi_{ \vert 011\rangle}$.

To conclude this section, we have demonstrated that an originally negative $J_\text{se}$ at $\varepsilon=0$ switches to a positive one at $\varepsilon\gg0$. We further deduce that the sign-switching of $J_\text{se}$ can be ascribed to the increase of the magnitude of the four-tunneling process term $\phi_{\vert S(002)\rangle}$ and the two-tunneling process term $\delta \phi_{\vert 011\rangle }$ at $\varepsilon\gg0$.

\section{Relevance to other works}\label{sec:compareOther}
Superexchange interaction between two delocalized electrons in a TQD device has been theoretically evaluated using Hubbard model in Ref.~\onlinecite{Rancic.17}. However, since Hubbard model only includes the on-site Coulomb interaction while excluding other types of Coulomb interaction, some subtle yet important Coulomb interaction terms might be overlooked. In fact, in contrast to the full CI results in Fig.~\ref{fig:staDiaandJ}, in the detuning region where the ground charge state is (101), the sign-switching and sweet spots of $J_\text{se}$ are \textit{not} found \cite{Rancic.17}. We focus on the case where the ground charge state is (101) since it is of experimental interests \cite{Malinowski.19,Chan.21,Qiao.21}. We further confirm that the inclusion of the inter-dot Coulomb interaction in the extended Hubbard model \cite{Kotlyar.98} \textit{does not} lead to the sign-switching of $J_\text{se}$ in the (101) region. This reaffirms that the sign-switching of $J_\text{se}$ is a result of the interplay between the tunneling term $J_\text{nn}\times \xi_{t,\text{se}}^2$ and the long-distance Coulomb exchange term $J^e_\text{L,R}$, as discussed in Sec.~\ref{subsubsec:hbLx0m}.

Experimentally, coherent coupling between two delocalized electrons in a TQD devices has been demonstrated \cite{Baart.17,Sanchez.14.2}. In particular, Ref.~\onlinecite{Baart.17} has demonstrated spin exchange between two distant spins whose superexchange energy is positive, i.e.~$J_\text{se}>0$. However, the superexchange interaction has only been demonstrated at two detuning points, as discussed in the second paragraph of Sec.~\ref{sec:JseVSlrmD}. In contrast, our work takes a step further to show an extensive study of $J_\text{se}$ in a larger range of detuning values, which reveals a richer behavior of $J_\text{se}$, cf.~Fig.~\ref{fig:staDiaandJ}.

\section{Conclusion and discussion}\label{sec:conclusion}
We have shown, using full CI calculations, that the superexchange interaction between two delocalized electrons in a TQD device exhibits a nontrivial behavior with respect to the control parameters. When $\varepsilon=0$, a symmetric TQD device with a larger confinement strength and shorter inter-dot distance yields a $J_\text{se}$ curve that switches from a positive value at small $\Delta$ to a negative value at large $\Delta$. Also, the changes between positive and negative values of $J_\text{se}$ are not monotonic, giving rise to a sweet spot with respective to $\Delta$ and $\varepsilon$. Enabled by the crossing from a positive $J_\text{se}$ to a negative $J_\text{se}$, the existence of the zero value $J_\text{se}$ is important to realize a \textit{true ``turn off"} of an exchange gate. Comparison between the numerical results of $J_\text{se}$ for $\varepsilon=0$ to the analytical expression of $J_\text{se}$ (Eq.~\eqref{eq:JseExp1Simpl}) indicates that the non-monotonic behavior of $J_\text{se}$ can be attributed to the interplay between the virtual nearest-neighbor exchange $J_\text{nn}$ and the higher-order tunneling induced term $\xi_{t,\text{se}}$. Also, the sign-switching of $J_\text{se}$ is found to be related to the competition between the tunneling term $J_\text{nn} \xi_{t,\text{se}}^2$ and the long-distance Coulomb exchange term $2J^e_\text{L,R}$. 

When $\varepsilon>0$, we have demonstrated that the magnitude of $J_\text{se}$ as a function of $\varepsilon$ is largely influenced by the tunneling term $J_\text{nn} \xi_{t,\text{se}}^2$, whose magnitude increases for a larger $\varepsilon$. At a larger $\varepsilon$, a positive $J_\text{se}$ exhibits a larger magnitude while a negative $J_\text{se}$ exhibits a smaller magnitude. The contrasting behavior of $\left|J_\text{se}\right|$ between opposite signs of $J_\text{se}$ stems from the negative sign of the long-distance Coulomb exchange term $-2J^e_\text{L,R}$. 

Entering the detuning regime where $\varepsilon\gg0$, we have found that an originally negative $J_\text{se}$ at $\varepsilon=0$ switches to a positive one at large $\varepsilon$. We have deduced that the sign-switching of $J_\text{se}$ can be ascribed to two reasons: (1) The enhanced admixture between the ground singlet, $\vert S(001)\rangle$, and the first excited singlet, $\vert S(002)\rangle$. This admixture arises from a four-tunneling process; (2) A larger admixture between the ground singlet, $\vert S(001)\rangle$, and the excited singlet, $\vert S(011)\rangle$, compared to the same type of admixture between triplets. This admixture arises from a two-tunneling process. The sign-switching observed for the values of $J_\text{se}$ suggests that this setup may serve as a quantum simulator when either an antiferromagnetic interaction, ferromagnetic interaction or both are required \cite{Kim.11,Islam.11,Islam.13,Garttner.17}.


Our results have shown that, depending on the quantum-dot and control parameters, the superexchange interaction mediated by an empty quantum dot in a TQD device exhibit a non-trivial behavior which ranges from positive to negative values. The richness of the properties of $J_\text{se}$ may be harnessed to enhance the capabilities of spin qubits in quantum dots.
	
	\section*{Acknowledgements} 
	G.X.C. and X.W. are supported by the Key-Area Research and Development Program of GuangDong Province  (Grant No.~2018B030326001), the National Natural Science Foundation of China (Grant No.~11874312), the Research Grants Council of Hong Kong (Grant No.~11303617), and the Guangdong Innovative and Entrepreneurial Research Team Program (Grant No.~2016ZT06D348). P.H. acknowledges supports by the National Natural Science Foundation of China (No. 11904157), Shenzhen Science and Technology Program (No. KQTD20200820113010023), and Guangdong Provincial Key Laboratory (No. 2019B121203002). The calculations involved in this work are mostly performed on the Tianhe-2 supercomputer at the National Supercomputer Center in Guangzhou, China.


\bibliographystyle{apsrev4-1}
\bibliography{SE2E3Dots}

\begin{thebibliography}{96}
\bibitem [{Jse()}]{JseDef}%
  \BibitemOpen
  \href@noop {} {}\bibinfo {note} {For two delocalized antiparallel spins in a
  TQD device with an empty mediator, superexchange interaction is demonstrated
  experimentally in the detuning regime where the charge configuration of the
  logical singlet and triplet states are (101) \cite{Baart.17}. This definition
  of superexchange interaction is also adopted in other experiements
  \cite{Malinowski.19,Chan.21,Qiao.21}. For example,
  Ref.~\onlinecite{Malinowski.19} has demonstrated superexchange interaction
  between two delocalized spins in a TQD device with a multielectron mediator
  in the detuning regime where the ground charge state is $(1N1)$ for $N$
  ranges from 50 to 100. Similarly, in a linear-quadrupole-quantum-dot device,
  Ref.~\onlinecite{Qiao.21} has demonstrated superexchange interaction between
  spins in two outmost dots in the detuning regime where the ground charge
  state is $(1111)$.}\BibitemShut {Stop}%
\end{thebibliography}

\setcounter{section}{0}
\setcounter{secnumdepth}{3}  
\setcounter{equation}{0}
\setcounter{figure}{0}
\setcounter{table}{0}
\renewcommand{\theequation}{S-\arabic{equation}}
\renewcommand{\thefigure}{S\arabic{figure}}
\renewcommand{\thetable}{S-\Roman{table}}
\renewcommand\figurename{Supplementary Figure}
\renewcommand\tablename{Supplementary Table}
\newcommand\Scitetwo[2]{[S\citealp{#1}, S\citealp{#2}]}
\newcommand\citeScite[2]{[\citealp{#1}, S\citealp{#2}]}

\newcommand\Scite[1]{[S\citealp{#1}]}
\newcommand\SciteOnline[1]{S\citealp{#1}}
\makeatletter \renewcommand\@biblabel[1]{[S#1]} \makeatother

\onecolumngrid

\begin{center}
\textbf{\large Supplementary Material for ``Universal control of superexchange in linear triple quantum dots with an empty mediator''	}
\end{center}

In this Supplemental Material, we provide necessary details complementary to results shown in the main text.

\section{full CI calculations}\label{sec:fullCI}
\subsection{Single-particle basis}
In this work, we use the solutions of the quadratic potential wells to approximate the single-particle basis. We assume the confining potential yields a quadratic in-plane potential for
electrons in a lateral gate-defined confinement of a quantum-dot (QD)
\begin{equation}
	V_{\mathbf{R}_0}(x,y)=\frac{1}{2}m^* \omega_0 \left[(x-x_0)^2+y^2\right],
\end{equation}
where the vector $\mathbf{R}_0=(x_0,0)$ is the position of potential minimum. The solutions to the single-particle Hamiltonian, ${(-i\hbar \nabla_j+e \mathbf{A}/c)^2}/{2m^*}+V(\mathbf{r})$, are the Fock-Darwin (F-D) states centered at the minimum of the potential well. The explicit expressions of F-D states are given in Appendix A in Ref.~\onlinecite{Barnes.11}. We denote the $j$-th lowest orthonormalized F-D states centered at each potential minimum in a TQD as $\Phi_{\text{L}j}$, $\Phi_{\text{M}j}$ and $\Phi_{\text{R}j}$ for the left, middle and right dots, respectively, cf.~Fig.~1 in the main text. The bare Fock-Darwin states centered at different well minima are not orthogonal and are further orthonormalized to construct orthonormal multielectron Slater determinants, cf.~Appendix B in Ref.~\onlinecite{Barnes.11}.
\subsection{Two-electron Slater determinant}
Since we are interested in the system with $S_z=0$, a generic two-electron Slater determinant can be written as
\begin{equation}
	\vert \Psi_j\rangle=\left\vert \uparrow_{\Phi_k} \downarrow_{\Phi_l}\right\rangle =\left|\begin {array} {cc}
	\Phi_k \left(\mathbf{r}_1\right)\uparrow \left(\omega_1\right)& \Phi_l \left(\mathbf{r}_1\right)\downarrow \left(\omega_1\right)  \\
	\Phi_k \left(\mathbf{r}_2\right)\uparrow \left(\omega_2\right)& \Phi_l \left(\mathbf{r}_2\right)\downarrow \left(\omega_2\right)  \\
	\end {array}
	\right|,
\end{equation}
where $\mathbf{r}_m$ and $\omega_m$ are the position vector and spin variable of the $m$-th electron, respectively, and $\Phi_{k}\in\{\Phi_{\text{L}1},\Phi_{\text{L}2},\cdots,\Phi_{\text{M}1},\Phi_{\text{M}2},\cdots,\Phi_{\text{R}1},\Phi_{\text{R}2},\cdots\}$. In this work, we consider all possible Slater determinants for a given set of F-D states.

\subsection{Matrix elements of $H$ and values of $J_\text{se}$}
The matrix form of the two-electron Hamiltonian $H$ (cf.~Eq.~1 in the main text), expanded in the bases of the Slater determinants, is
\begin{equation}\label{eq:HSupp}
	H=\sum_{j,k}^{N_{\Psi}} (H_{j,k}^h + H_{j,k}^C) \vert \Psi_j\rangle \langle \Psi_k\vert,
\end{equation}
where $N_{\Psi}$ is the number of the Slater determinants, $H_{j,k}^h=\langle \Psi_j \vert h \vert \Psi_k \rangle$ is the matrix element of the single-particle Hamiltonian, and $H_{j,k}^C=\langle \Psi_j \vert H_C \vert \Psi_k \rangle$ is the matrix element of the Coulomb interaction. The matrix elements $H_{j,k}^h$ $\left(H_{j,k}^C\right)$ can be expanded as sums of the integrals $ \int \Phi_m^* \left(\mathbf{r}\right) h \Phi_n \left(\mathbf{r}\right) d\mathbf{r}$ $\left[ \int \Phi_m^*(\mathbf{r}_1) \Phi_n^*(\mathbf{r}_2) H_C \Phi_o (\mathbf{r}_1) \Phi_p (\mathbf{r}_2) d\mathbf{r}_1 d\mathbf{r}_2 \right]$ according to Slater-Condon rules \cite{Szabo.96}. Details of the evaluation of the integrals are provided in Appendix B in Ref.~\onlinecite{Barnes.11}. 

The values of $J_\text{se}$ are then obtained by diagonalizing $H$ in Eq.~\eqref{eq:HSupp} and taking the energy difference between the lowest singlet and triplet states.

\section{The generalized form of Hubbard model}
The explicit form of the generalized Hubbard model is
\begin{equation}\label{eq:HubbardModel2ndQuanSupp}
	\begin{split}
		H_\text{Hubbard} &= \sum_{j \alpha \sigma}\mu_{j \alpha } c^\dagger_{j \alpha \sigma} c_{j \alpha \sigma}
		+\sum_{j \neq j',\alpha,\alpha',\sigma} t_{j \alpha ,j' \alpha'}  c^\dagger_{j \alpha \sigma} c_{j' \alpha' \sigma}
		+\sum_{j\alpha} U_{j \alpha} n_{j \alpha\downarrow} n_{j\alpha \uparrow}
		+\sum_{\sigma \sigma'}\sum_{j\alpha \neq j'\alpha'} U'_{j \alpha,j' \alpha'} n_{j \alpha \sigma} n_{j' \alpha '\sigma'} 
		\\
		&\quad-\sum_{\sigma\neq\sigma'}\sum_{j \alpha \neq j' \alpha'}J^e_{j \alpha, j' \alpha'}c^\dagger_{j \alpha \sigma}c^\dagger_{j' \alpha' \sigma'}c_{j' \alpha' \sigma}c_{j \alpha \sigma}
		-\sum_{\sigma\neq\sigma'}\sum_{j \alpha \neq j' \alpha'}J^p_{j \alpha, j' \alpha'}c^\dagger_{j \alpha \sigma}c^\dagger_{j \alpha \sigma'}c_{j' \alpha' \sigma}c_{j' \alpha' \sigma'}
		\\
		&\quad +\sum_{\sigma\neq\sigma'}\sum_{j, j' \alpha' \neq j'' \alpha''}J^t_{j \alpha(j' \alpha',j'' \alpha'')}c^\dagger_{j \alpha\sigma}c_{j \alpha\sigma}c^\dagger_{j' \alpha' \sigma'}c_{j'' \alpha'' \sigma'}
		,
	\end{split}
\end{equation}
where
\begin{equation}\label{eq:HubbardParameters}
	\begin{split}
		\mu_{j\alpha} &= \int d\mathbf{r}_{j} \Phi_{j\alpha}(\mathbf{r}) h  \Phi_{j\alpha}(\mathbf{r}),
		\\
		t_{j \alpha,j' \alpha'} &= \int d\mathbf{r}_j \Phi_{j \alpha}(\mathbf{r}) h  \Phi_{j' \alpha'}(\mathbf{r}),\\
		U_{j\alpha}&=\int d\mathbf{r}_1 d\mathbf{r}_2 \Phi^*_{j\alpha}(\mathbf{r}_1) \Phi^*_{j \alpha}(\mathbf{r}_2) H_C  \Phi_{j \alpha}(\mathbf{r}_1) \Phi_{j \alpha}(\mathbf{r}_2),
		\\
		U'_{j \alpha,j' \alpha'}&=\int d\mathbf{r}_1 d\mathbf{r}_2 \Phi^*_{j \alpha}(\mathbf{r}_1) \Phi^*_{j' \alpha'}(\mathbf{r}_2) H_C \Phi_{j \alpha}(\mathbf{r}_1) \Phi_{j' \alpha'}(\mathbf{r}_2),
		\\
		J^e_{j \alpha,j' \alpha'}&=\int d\mathbf{r}_1 d\mathbf{r}_2 \Phi^*_{j \alpha}(\mathbf{r}_1) \Phi^*_{j' \alpha'}(\mathbf{r}_2) H_C \Phi_{j' \alpha'}(\mathbf{r}_1) \Phi_{j \alpha}(\mathbf{r}_2),
		\\
		J^p_{j \alpha,j' \alpha'}&=\int d\mathbf{r}_1 d\mathbf{r}_2 \Phi^*_{j \alpha}(\mathbf{r}_1) \Phi^*_{j \alpha}(\mathbf{r}_2) H_C 
		\Phi_{j' \alpha'}(\mathbf{r}_1) \Phi_{j' \alpha'}(\mathbf{r}_2),
		\\
		J^t_{j \alpha(j' \alpha',j'' \alpha'')}&=\int d\mathbf{r}_1 d\mathbf{r}_2 \Phi^*_{j \alpha}(\mathbf{r}_1) \Phi^*_{j' \alpha'}(\mathbf{r}_2) H_C 
		\Phi_{j \alpha}(\mathbf{r}_1) \Phi_{j'' \alpha''}(\mathbf{r}_2),
	\end{split}
\end{equation}
In Eq.~\eqref{eq:HubbardModel2ndQuanSupp}, $\{j \alpha,k \alpha,l \alpha\}\in\{\text{L}, \text{M}, \text{R}\}$ refer to the $\alpha$-th lowest orbitals in left (L), middle (M) and right (R) dots, respectively, while  $\sigma,\sigma'$ refer to the spins $\{\sigma,\sigma'\}\in\{\uparrow,\downarrow\}$. $\mu_{j \alpha\sigma}$ is the chemical potential of an electron with spin $\sigma$ occupying the orbital $j \alpha$, $t_{j \alpha,j' \alpha'\sigma}$ is the tunneling energy between the orbitals $j \alpha$ and $j' \alpha'$, $U_{j \alpha}$ is the on-site Coulomb interaction in orbital $j \alpha$, $U'_{j \alpha,j' \alpha'}$ is the inter-orbital Coulomb interaction between the orbitals $j \alpha$ and $j' \alpha'$, $J^e_{j \alpha,j' \alpha'}$ is the Coulomb exchange interaction between the orbitals $j \alpha$ and $j' \alpha'$. In Eq.~4 in the main text, we have dropped the terms $J^p$ and $J^t$ which are negligible.

	\section{Analytical expressions of superexchange energy $J_\mathrm{se}$ }\label{sec:JseAna}
	\subsection{$\Delta> 0$, $\varepsilon\geq 0$, $E_{\vert S(002)\rangle} > E_{\vert S(011)\rangle}$ in (101) region}\label{subsec:JseDLRNonzero}
	For $\Delta>0$, we derive here the expression of $J_\text{se}$ by considering the regime where $\varepsilon\geq0$. The lowest energy subspace are spanned by $\vert S(101)\rangle$, $\vert S(011)\rangle$ for singlets and $\vert T(101)\rangle$, $\vert T(011) \rangle$ for triplets, cf.~Fig.~2(a) in the main text. Using the generic Hubbard model (Eq.~4 in the main text), the Hamiltonian blocks written in the bases above for singlets and triplets are
	
	\begin{subequations}\label{eq:HamEffSJMod}
		\begin{align}
			\begin{split}\label{eq:singHam1Mod}
				H_S=\left(
				\begin{array}{cc}
					E_{\vert S(101)\rangle} & t \\
					t & E_{\vert S(011)\rangle}-t \xi_{t,\text{nn}} \\
				\end{array}
				\right),
			\end{split}\\
			\begin{split}\label{eq:tripHamMod}
				H_T=\left(
				\begin{array}{cc}
					E_{\vert T(101)}& t \\
					t & E_{\vert T(011)\rangle} \\
				\end{array}
				\right),
			\end{split}
		\end{align}
	\end{subequations}
	respectively, where $E_{\vert S(101)\rangle}=J^e_\text{L,R}+U'_\text{L,R}$, $E_{\vert S(011)\rangle}=\left(U'_\text{M,R}+\Delta-\varepsilon\right)+J^e_\text{M,R}$, $E_{T(101)\rangle}=-J^e_\text{L,R}+U'_\text{L,R}$, $E_{T(011)\rangle}=\left(U'_\text{M,R}+\Delta-\varepsilon\right)-J^e_\text{M,R}$,  $t=t_\text{L,M}=t_\text{M,R}$. An additional term $t \xi_{t,\text{nn}}$ is obtained by diagonalizing the Hamiltonian blocks spanned by the first and second excited singlet states, i.e.~$\vert S(011)\rangle$ and $\vert S(002)\rangle$, respectively. Taking the lowest eigenvalues of $H_S$ and $H_T$, we obtain an analytical expression of $J_\text{se}$, as given in Eq.~5 in the main text. 
	
	\subsection{$\Delta>0$, $\varepsilon \gg 0$, $E_{\vert S(011)\rangle} >E_{\vert S(002)\rangle} $ in (101) region} \label{subsec:JseDLRLarge}
	In the regime where $\Delta>0$ and $\varepsilon\gg 0$ such that $E_{\vert S(011)\rangle} > E_{\vert S(002)\rangle}$ in (101) region, the lowest energy bases are $\vert S(101)\rangle$, $\vert S(002)\rangle$, $\vert S(011)\rangle$ for singlets and $\vert T(101)\rangle$, $\vert T(011)\rangle$ for triplets, cf.~Fig.~2(b) in the main text. Using the generic Hubbard model (Eq.~4 in the main text), the Hamiltonian blocks written in the bases above for singlets and triplets are
	
	\begin{subequations}\label{eq:HDLRLarge}
		\begin{align}
			H_S&=\left(
			\begin{array}{ccc}
				E_{\vert S(101)\rangle} & 0 &  t \\
				0 & E_{\vert S(002)\rangle} & t \\
				t & t  & E_{\vert S(011)\rangle}\\
			\end{array}
			\right),
			\\
			H_T&=\left(
			\begin{array}{cc}
				E_{\vert T(101)\rangle} & t  \\
				t & E_{\vert T(011)\rangle}
			\end{array}
			\right),\label{eq:HT}
		\end{align}
	\end{subequations}
	respectively, where $E_{\vert S(101)\rangle}=J^e_\text{L,R}+U'_\text{L,R}$, $E_{\vert S(002)\rangle}=U_\text{R}-2\varepsilon$, $E_{\vert S(011)\rangle}=J^e_\text{M,R}+U'_\text{M,R}+\Delta-\varepsilon$, $E_{\vert T(101)\rangle}=-J^e_\text{L,R}+U'_\text{L,R}$, and $E_{\vert T(011)\rangle}=-J^e_\text{M,R}+U'_\text{M,R}+\Delta-\varepsilon$.
	The eigenvalue of ground singlet state, $E_{\vert S(101)\rangle'}$, is obtained by taking the minimum of the diagonal terms of ${U^{(2)}}^\dagger{U^{(1)}}^\dagger H_S U^{(1)} U^{(2)}$, where $U^{(1)}$ diagonalizes the subspace spanned by the first and second excited singlet states, i.e.~$E_{\vert S(011)\rangle}$ and $E_{\vert S(002)\rangle}$, respectively, while $U^{(2)}$ is the S-W transformation up to $\mathcal{O}\left[t^2/(E_{\vert\eta\rangle_2}-E_{\vert\eta\rangle_1})\right]$. Taking the difference of the lowest eigenvalue of $H_T$ and $E_{\vert S(101)\rangle'}$, we obtain the expression of $J_\text{se}$, as shown in Eq.~7 in the main text. The procedure above, in particular the transformation ${U^{(2)}}^\dagger{U^{(1)}}^\dagger H_S U^{(1)} U^{(2)}$, allows us to relate $J_\text{se}$ with the admixture probabilities of excited states in the logical eigenstates, $\phi$, as shown in Eq.~8 in the main text.
	
	\section{Extraction of Hubbard parameters from CI results}
	\begin{figure}
		\includegraphics[width=0.7\columnwidth]{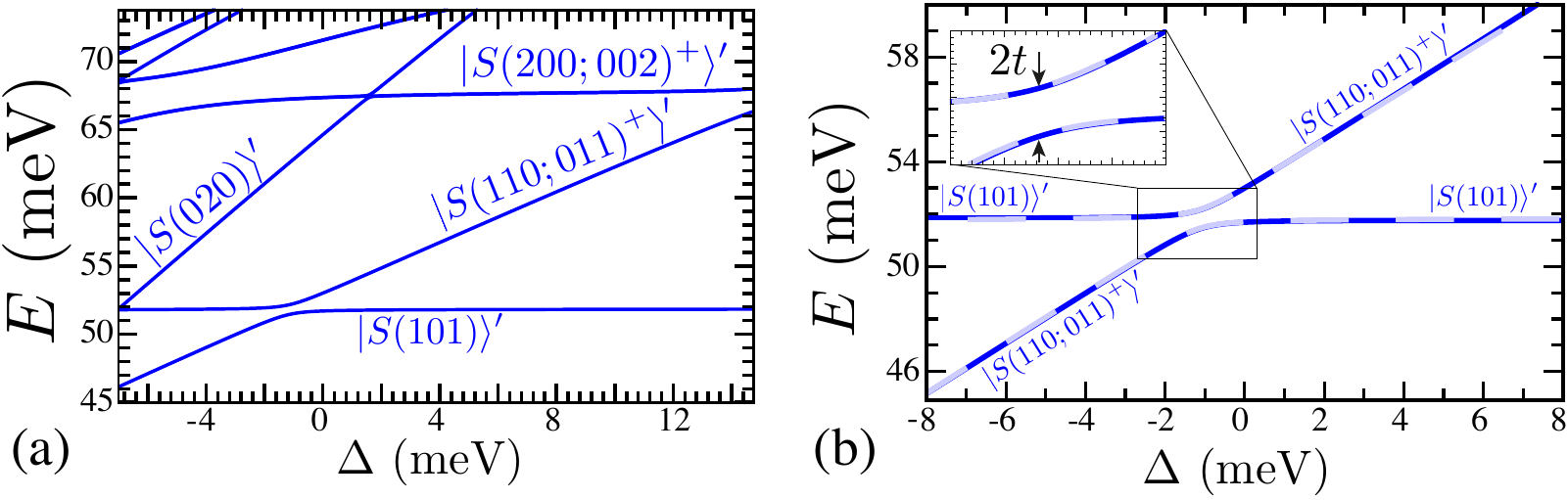}
		\caption{(a) CI results for the eigenvalues of singlet states obtained. (b) Correspondence between the CI results (solid lines) and the generic Hubbard Model (dashed lines). The inset shows the zoom-in part near the anticrossing between $\vert S(101)\rangle'$ and $\vert S(110;011)^+\rangle'$. The parameters are $\hbar \omega_0=25$ meV, $x_0=30$ nm, and $\varepsilon=0$. The eigenvalues shown in (a) are equivalent to those in Fig.~4(c) in the main text.}
		\label{fig:Sevals}
	\end{figure}
	Comprehending the behavior of $J_\text{se}$ from CI results requires mapping the lowest few eigenstates to an effective Hamiltonian given by the generic Hubbard model, cf.~Eqs.~\eqref{eq:HamEffSJMod} and \eqref{eq:HDLRLarge}. Here, we use the eigenvalues given in Fig.~\ref{fig:Sevals} as an example to illustrate the mapping and the method applies to the discussions of $J_\text{se}$ in other parameter regimes in the main text.

	The first step starts from identifying the eigenstates from CI results. At $\Delta=4$ meV and $\varepsilon=0$, inspecting the eigenstate that gives the eigenvalue $\approx52$ meV tells us that it is an eigenstate formed by one electron occupying the left dot while another electron occupying the right dot. Therefore, we label it as $\vert S(101)\rangle'$. We then proceed to identify $\vert S(110;011)^+\rangle'$ and $\vert S(200;002)^+\rangle'$, cf.~Fig.~\ref{fig:Sevals}(a).

	
	From CI results [Fig.~\ref{fig:Sevals}(a)], at $\Delta=4$ meV, we have $E_{\vert S(101)\rangle}=51.8$ meV, $E_{\vert S(011)\rangle}=56.7$ meV, and $E_{\vert S(002)\rangle}=67.6$ meV (the equalities, $E_{\vert S(110;011)^+\rangle}=E_{\vert S(011)\rangle}$ and $E_{\vert S(200;002)^+\rangle}=E_{\vert S(002)\rangle}$, hold for $\varepsilon=0$). The matching between the states $\vert \eta \rangle$ and the eigenvalues from CI results for other values of $\Delta$ gives us the values of $E_{\vert \eta \rangle}$ v.s.~$\Delta$.

	To obtain the value of inter-dot tunneling $t$, we input the values of $E_{\vert \eta \rangle}$ obtained above into Eq.~\eqref{eq:HamEffSJMod} and reproduce the eigenvalues for a selected value of $t$, cf.~Fig.~\ref{fig:Sevals}(b). The value of $t$ is then determined if the reproduced eigenvalues match with those from CI results, especially around the region where the anticrossing is found (the inset in Fig.~\ref{fig:Sevals}(b)). For the example shown here, $t=255$ $\mu$eV.

	Finally, the values of $\xi_{t,l}$ for $l\in\{\text{se},\text{nn}\}$ in Eq.~7 in the main text are evaluated using the values of $E_{\vert \eta \rangle}$ and $t$ obtained above. These values are shown in Figs.~4(b) and (f) in the main text. The values of $\xi$ shown in Figs.~5(d) and (h) are also obtained using the same procedure above.

	\section{Probabilities of excited states in the logical states, $\phi$}
	In the regime of $\varepsilon\gg0$, the behavior of $J_\text{se}$ is more easily understood using the probabilities of excited states in the logical states, denoted as $\phi$, cf.~Eq.~8 in the main text. For the logical singlet state, to obtain the values of $\phi_{\vert S(N_\text{L}N_\text{M}N_\text{R})\rangle}$, we first take the explicit from of the logical eigenstate ($\vert S(101)\rangle'$) from CI results, which are expanded in the bases of all the Slater determinants. We then distinguish the total probabilities of the Slater determinants with different charge configurations $(N_\text{L}N_\text{M}N_\text{R})$. The same goes for the logical triplet state.

	Here, we use the CI results for $\hbar \omega_0=7.28$ meV, $x_0=50$ nm as an example. At $\Delta=4.5$ meV and $\varepsilon=3.4$ meV, the lowest singlet state is
	\begin{equation}
		\begin{split}
			\vert S(101)\rangle'&=0.697 \left(\left\vert \uparrow_{\Phi_\text{L1}}\downarrow_{\Phi_\text{R1}}\right\rangle+\left\vert \uparrow_{\Phi_\text{R1}}\downarrow_{\Phi_\text{L1}}\right\rangle\right)+\cdots
			\\
			&\quad +0.087  \left(\left\vert \uparrow_{\Phi_\text{R1}}\downarrow_{\Phi_\text{M1}}\right\rangle+\left\vert \uparrow_{\Phi_\text{M1}}\downarrow_{\Phi_\text{R1}}\right\rangle\right)
			-0.017  \left(\left\vert \uparrow_{\Phi_\text{R1}}\downarrow_{\Phi_\text{M2}}\right\rangle+\left\vert \uparrow_{\Phi_\text{M2}}\downarrow_{\Phi_\text{R1}}\right\rangle\right)
			+\cdots
			\\
			&\quad +0.047  \left(\left\vert \uparrow_{\Phi_\text{R1}}\downarrow_{\Phi_\text{R1}}\right\rangle+\left\vert \uparrow_{\Phi_\text{R1}}\downarrow_{\Phi_\text{R1}}\right\rangle\right)
			+0.016  \left(\left\vert \uparrow_{\Phi_\text{R1}}\downarrow_{\Phi_\text{R5}}\right\rangle+\left\vert \uparrow_{\Phi_\text{R5}}\downarrow_{\Phi_\text{R1}}\right\rangle\right)
			+\cdots
		\end{split}
	\end{equation}
	The first, second and third lines correspond to the Slater determinants with $(N_\text{L}N_\text{M}N_\text{R})=(101), (011),$ and $(002)$, respectively. $(\cdots)$ refers to other Slater determinants whose coefficients exhibit smaller magnitude compared to those explicitly shown. The total of the absolute square of the coefficients in the second (third) line gives $\phi_{\vert (011)\rangle}$ $\left[\phi_{\vert (002)\rangle}\right]$, as given in Figs.~6(c) and (d) in the main text.
\end{document}